\RequirePackage{fix-cm}

\documentclass[twocolumn,epjc3]{svjour3}  

\smartqed
\usepackage{comment}
\usepackage{amsmath}
\usepackage{multirow}
\usepackage{comment}
\usepackage{float}
\usepackage[justification=justified]{subcaption}
\usepackage[justification=justified, figurename=Fig.,tablename=Tab.,font=small,margin=0em]{caption}
\usepackage{graphicx}
\usepackage[dvipsnames]{xcolor} 
\usepackage{dcolumn}
\usepackage{bm}
\usepackage{hyperref}
\usepackage{xfrac}
\usepackage{cite}

\usepackage{threeparttable} 
\usepackage{booktabs} 
\usepackage[separate-uncertainty,retain-explicit-plus,per-mode=symbol,binary-units,parse-numbers=false]{siunitx}
\DeclareSIUnit\parsec{pc}
\usepackage{enumitem}
\usepackage{csquotes}
\usepackage{stmaryrd}
\usepackage{outlines} 

\newcommand\RevE[1]{\textcolor{black}{#1}}

\newcommand{\curlyL}{\mathcal{L}} 
\newcommand{\dquotes}[1]{``#1''} 
\newcommand{\tx}{\text}

\newcommand{\gevm}{\ensuremath{\mathrm{GeV}/c^2}}
\newcommand{\kev}{\ensuremath{\mathrm{keV}}}

\hypersetup{
    colorlinks=true,
    linkcolor=blue,
    citecolor=blue,
    filecolor=magenta,
    urlcolor=blue,
}

\begin{document}
\title{{Recommended conventions for reporting results from direct dark matter searches}}

\author{D.~Baxter\thanksref{ad1} \and 
I.~M.~Bloch\thanksref{ad2} \and 
E.~Bodnia\thanksref{ad3} \and
X.~Chen\thanksref{ad4,ad5} \and
J.~Conrad\thanksref{ad6} \and 
P.~Di~Gangi\thanksref{ad7} \and 
J.~E.~Y.~Dobson\thanksref{ad8} \and 
D. Durnford\thanksref{ad9} \and 
S.~J.~Haselschwardt\thanksref{ad10} \and
A.~Kaboth\thanksref{ad11,ad12}  \and 
R.~F.~Lang\thanksref{ad13} \and 
Q.~Lin\thanksref{ad14} \and 
W.~H.~Lippincott\thanksref{ad3} \and 
J.~Liu\thanksref{ad4, ad5, ad15} \and 
A.~Manalaysay\thanksref{ad10} \and 
C.~McCabe\thanksref{ad16}  \and 
K.~D.~Mor\aa\thanksref{ad17} \and 
D.~Naim\thanksref{ad18} \and 
R.~Neilson\thanksref{ad19} \and 
I.~Olcina\thanksref{ad10, ad20} \and 
M.-C.~Piro\thanksref{ad9} \and 
M.~Selvi\thanksref{ad7} \and 
B.~von~Krosigk\thanksref{ad22} \and 
S.~Westerdale\thanksref{ad23} \and 
Y.~Yang\thanksref{ad4} \and 
N.~Zhou\thanksref{ad4}
}

\institute{Kavli Institute for Cosmological Physics and Enrico Fermi Institute, University of Chicago, Chicago, IL 60637 USA \label{ad1} \and 
School of Physics and Astronomy, Tel-Aviv University, Tel-Aviv 69978, Israel \label{ad2} \and
University of California, Santa Barbara, Department of Physics, Santa Barbara, CA 93106, USA \label{ad3} \and
INPAC and School of Physics and Astronomy, Shanghai Jiao Tong University, MOE Key Lab for Particle Physics, Astrophysics and Cosmology, Shanghai Key Laboratory for Particle Physics and Cosmology, Shanghai 200240, China \label{ad4} \and
Shanghai Jiao Tong University Sichuan Research Institute, Chengdu 610213, China \label{ad5} \and
Oskar Klein Centre, Department of Physics, Stockholm University, AlbaNova, Stockholm SE-10691, Sweden \label{ad6} \and
Department of Physics and Astronomy, University of Bologna and INFN-Bologna, 40126 Bologna, Italy \label{ad7} \and
University College London, Department of Physics and Astronomy, London WC1E 6BT, UK \label{ad8} \and
Department of Physics, University of Alberta, Edmonton, Alberta, T6G 2R3, Canada \label{ad9} \and
Lawrence Berkeley National Laboratory (LBNL), Berkeley, CA 94720, USA \label{ad10} \and 
STFC Rutherford Appleton Laboratory (RAL), Didcot, OX11 0QX, UK \label{ad11} \and
Royal Holloway, University of London, Department of Physics, Egham, TW20 0EX, UK \label{ad12} \and
Department of Physics and Astronomy, Purdue University, West Lafayette, IN 47907, USA \label{ad13} \and
Department of Modern Physics, University of Science and Technology of China, Hefei, Anhui 230026, China \label{ad14} \and
Tsung-Dao Lee Institute, Shanghai 200240, China \label{ad15} \and
King’s College London, Department of Physics, London, WC2R 2LS, UK \label{ad16} \and
Columbia University, Columbia Astrophysics Lab, New York, NY 10027, USA \label{ad17} \and
University of California, Davis, Department of Physics, Davis, CA 95616, USA \label{ad18}\and
Drexel University, Department of Physics, Philadelphia, PA 19104, USA \label{ad19}\and
University of California, Berkeley, Department of Physics, Berkeley, CA 94720, USA \label{ad20}\and
Institut  f{\"u}r  Experimentalphysik,  Universit{\"a}t  Hamburg,  22761  Hamburg,  Germany \label{ad22}\and
INFN Cagliari, Cagliari 09042, Italy \label{ad23}
}

\date{Received: date / Accepted: date}

\maketitle
\begin{abstract}
The field of dark matter detection is a highly visible and highly competitive one. 
In this paper, we propose recommendations for presenting dark matter direct detection results particularly suited for weak-scale dark matter searches, although we believe the spirit of the recommendations can apply more broadly to searches for other dark matter candidates, such as very light dark matter or axions.
To translate experimental data into a final published result, direct detection collaborations must make a series of choices in their analysis, ranging from how to model astrophysical parameters to how to make statistical inferences based on observed data. While many collaborations follow a standard set of recommendations in some areas, for example the expected flux of dark matter particles (to a large degree based on a paper from Lewin and Smith in 1995), in other areas, particularly in statistical inference, they have taken different approaches, often from result to result by the same collaboration. 
We set out a number of recommendations on how to apply the now commonly used Profile Likelihood Ratio method to direct detection data.  In addition, updated recommendations for the Standard Halo Model astrophysical parameters and relevant neutrino fluxes are provided.  The authors of this note include members of the DAMIC, DarkSide, DARWIN, DEAP, LZ, NEWS-G, PandaX, PICO, SBC, SENSEI, SuperCDMS, and XENON collaborations, and these collaborations provided input to the recommendations laid out here. 
Wide-spread adoption of these recommendations will make it easier to compare and combine future dark matter results. 
\end{abstract}

\section{Introduction and Purpose of this Paper}
The nature of dark matter (DM) is one of the highest-priority topics in high energy particle physics. Many collaborations around the world are building exquisitely sensitive detectors to search for dark matter particles, often in direct competition with each other, and in the future, collaborations may wish to combine data from complementary targets to draw even stronger conclusions about dark matter models, especially in light of neutrino backgrounds~\cite{Ruppin:2014bra} and model uncertainties~\cite{buch_implications_2020}. 

In going from data to a final dark matter result, or even in projecting the potential sensitivity of a proposed experiment, direct detection collaborations make a series of choices, ranging from how to model the dark matter halo in the Milky Way to which test statistic to use to perform statistical inference.   Different approaches can lead to significant differences in the interpretation of a result even if the underlying data are the same, complicating comparisons and combinations of results. 
In a recent example, the LUX collaboration deployed a power constrained limit~\cite{PCL} (discussed in Sec.~\ref{sec:limitedpower}) for their dark matter limits \cite{lux_first,lux_collaboration_results_2017}, but chose a different power threshold in the two results; making the same choice in Ref.~\cite{lux_first} as in  Ref.~\cite{lux_collaboration_results_2017} would have changed the resulting limit by a factor of $\sim$2. Similarly, the XENON1T collaboration presented a first result by approximating their likelihood ratio  with an asymptotic distribution~\cite{Aprile:2017iyp}, an approximation that  led incorrectly to a $\sim$50\% more sensitive result. For their second science run, XENON1T corrected this treatment~\cite{xenon1t_sr1}. 

Background modeling is another  area where collaborations make choices with potentially significant implications on inferred results. While many backgrounds are unique to each detector, there are some elements that are shared by all direct detection experiments, such as those induced by astrophysical neutrinos. To model solar or atmospheric neutrino backgrounds, collaborations rely on external data, with varying possible interpretations of the rates in dark matter detectors. As direct detection experiments increase in exposure, measurements of these astrophysical neutrino fluxes will be among the primary determinants of sensitivity~\cite{OHare:2020lva}.

Dating back to the paper of Lewin and Smith~\cite{lewin_review_1996},  dark matter collaborations have mostly (but not entirely) used similar assumptions about the phase-space distribution of dark matter. However, the community has not converged on a similar consensus regarding how to make statistical inferences from direct detection data. In this paper, we lay out recommendations for statistical methods aimed primarily at the Profile Likelihood Ratio (PLR) method, now commonly used in searches for weak-scale dark matter candidates, although some of these recommendations do apply more generally. We recognize that not all analyses lend themselves to the PLR and we hope that collaborations will follow the spirit of these recommendations when applicable. 
We take the opportunity to make updated recommendations for modeling the distribution of dark matter in our galaxy, as well as to discuss neutrino backgrounds that will be observed by many experiments in the near future.

This effort grew out of a Phystat-DM workshop~\cite{Phystat} held in Stockholm, Sweden, in August 2019, under the umbrella of the Phystat conference series. The authors of this note include members of the DAMIC, DarkSide, DARWIN, DEAP, LZ, NEWS-G, PandaX, PICO, SBC, SENSEI, SuperCDMS, and XENON collaborations, and these collaborations provided input to the recommendations laid out here. Our approach is similar in spirit to that of the ATLAS and CMS experiments in the period prior to the discovery of the Higgs, when the two collaborations agreed in advance on what statistical treatment to use in combining Higgs data sets~\cite{ATLAS:2011tau}, although we make different recommendations that we feel are more appropriate for our application.

In writing this white paper, we recognize the large influence of chance when analysing dark matter data; due to the low backgrounds, the expected statistical fluctuations for direct detection upper limits are around twice as large as those in the Higgs discovery~\cite{atlas_higgsdiscovery, Chatrchyan:2013lba}. Nevertheless, settling on common standards will enable more accurate comparisons of projections and results from different experiments and technologies as well as statistical combinations across collaborations.  If, as we hope will be the case, this work is used as a reference in future dark matter publications, the underlying works on which our recommendations are based should also be cited.
In addition to the specific recommendations given here, we suggest that collaborations continue to communicate with each other on these topics and adapt as necessary when new results are released. 

The paper is organized as follows: Section~\ref{sec:PLR_analyses} discusses Profile Likelihood Ratio analyses, Section~\ref{sec:AP} discusses astrophysical models, with Section~\ref{sec:Halo} focusing on the dark matter halo distribution, summarized in Table~\ref{tab:shmparams}, and Section~\ref{sec:AP_nu} focusing on astrophysical neutrinos, summarized in Table~\ref{tab:proposed_fluxes}.  An overall summary of our recommendations is provided in Section~\ref{sec:summary}.


\section{Profile Likelihood Ratio Analyses}
\label{sec:PLR_analyses}
Frequentist hypothesis testing has traditionally been the preferred method in direct dark matter searches to place constraints on regions of parameter space.  Our recommendations are developed for analyses deploying the profile likelihood ratio (PLR) method~\cite{Rolke:2004mj,CowanAsymptotic, PDG}, although some can be applied more generally.
Using a likelihood-based test statistic like the PLR has the advantage that  experimental uncertainties can conveniently be accounted for by parameterizing them as nuisance parameters. The PLR method has been described in great detail elsewhere, and we follow the discussion and notation of Ref.~\cite{CowanAsymptotic}. We strongly recommend readers to review Sec.~2 of Ref.~\cite{CowanAsymptotic} and Ref.~\cite{PDG} as we do not attempt to cover the subject fully here. 

For a set of parameters of interest, $\bm{\mu}$, and a collection of nuisance parameters, $\bm{\theta}$, the profile likelihood ratio is defined as 
\begin{equation}
  \lambda(\bm{\mu}) \equiv \frac{\curlyL(\bm{\mu},\hat{\hat{\bm{\theta}}})}{\curlyL(\hat{\bm{\mu}},\hat{\bm{\theta}})} \ ,
  \label{eq:plr}
\end{equation}
%
with $\curlyL$ as the likelihood function. The maximum of $\curlyL$ is found at $\hat{\bm{\mu}}$ and $\hat{\bm{\theta}}$, the maximum likelihood (ML) estimators of the parameters, while the maximum for a given $\bm{\mu}$ is found at $\hat{\hat{\bm{\theta}}}$. By construction, the parameter $\lambda(\bm{\mu})$ is constrained between 0 and 1, and values of $\lambda$ close to 1 are indicative of a good agreement between the data and the hypothesized value of $\bm{\mu}$.  

Direct dark matter searches most often take the hypothesis under test to be a signal model (generally the signal strength or cross section $\sigma$) at a single dark matter mass, $M$, and then 2D curves are constructed by computing significance and confidence intervals for each mass separately.
In this strategy, known as a ``raster scan'', only a single parameter of interest is constrained and therefore, $\bm{\mu} = \mu$ (as the name suggests, the procedure is typically repeated for different, fixed values of other signal parameters such as particle mass).  An alternative 2D approach would be to constrain $\sigma$ and $M$ at the same time. As discussed in Ref.~\cite{rasterscan}, the raster scan looks for the best region of $\sigma$ at each mass $M$ separately, while the 2D approach searches for a region around optimal values of $\sigma$ and $M$. For the reasons laid out in Ref.~\cite{rasterscan} and in keeping with convention to date, we advocate following the raster scan approach in most of what follows, but we return to this question in Sec.~\ref{contour}.

Given $\lambda({\mu})$, one can define
\begin{equation}
  t_{\mu} \equiv -2\log\lambda({{\mu}}) ,
  \label{eq:log_likelihood}
\end{equation}
which is distributed between 0 and infinity. As originally shown in Ref.~\cite{Wilks}, Wilks' theorem states that the distribution of $t_{\mu}$ approaches a chi-square distribution in the asymptotic limit of infinite data. Several conditions must be fulfilled for Wilks' theorem to hold, including that the true value of all parameters should be in the interior of the parameter space, that the sample should be sufficiently large, that no parameter may vary without the model changing, and that the hypotheses are nested~\cite{natrev_acms}.

The level of disagreement between the observed
data and the hypothesis under test (a given value of $\mu$) is usually quantified via the $p$-value. This corresponds to the probability for getting a value of $t_\mu$ for a given $\mu$ as large, or larger, than the one observed:
\begin{equation}
\begin{aligned}
  p_{{\mu}} &= P(t_{{\mu}} \ge t_{{\mu},\tx{obs}} | {\mu}) 
  &= \int\limits_{t_{\tx{obs}}}^{\infty} f(t_{{\mu}} | {\mu}) dt_{{\mu}} ,
\end{aligned}
  \label{eq:p_two_sided}
\end{equation}
where $f(t_{{\mu}} | {\mu})$ is the probability density function for $t_{{\mu}}$.


In the case of dark matter, the sought-after signal can only increase the data count
(i.e.~$\mu$ is defined strictly positive). One can modify Eq.~\ref{eq:log_likelihood} to become
\begin{equation}
  \tilde{t}_{\mu} =
  \begin{cases}
    -2 \log \frac{\curlyL(\mu,\hat{\hat{\bm{\theta}}})}{\curlyL(\hat{\mu},\hat{\bm{\theta}})} & \hat{\mu} \ge 0,\\
    -2 \log \frac{\curlyL(\mu,\hat{\hat{\bm{\theta}}})}{\curlyL(0,\hat{\bm{\theta}}(0))} & \hat{\mu} < 0 ,
  \end{cases}
  \label{eq:plr_two_sided_mu_positive}
\end{equation}
which takes into account that for $\hat{\mu}<0$, the maximum likelihood estimator will always be $\mu=0$. Note that here we follow the prescription in Ref.~\cite{CowanAsymptotic}, which treats $\hat{\mu}$ as an effective estimator that can take negative values even if the condition $\mu \ge 0$ is required by the physical model.




\subsection{Discovery}
\label{sec:discovery}

The primary objective for direct detection experiments is to search for the presence of new signal processes. In this case, the background-only null hypothesis, $H_0$, with $\mu=0$, is the crucial hypothesis to test. With signals expected to lead to an excess of events over the background, a special case of the test statistic in Eq.~\ref{eq:plr_two_sided_mu_positive} evaluated at $\mu=0$,  $\tilde{t}_0$ (also called $q_0$ in Ref.~\cite{CowanAsymptotic}),  should be used to assess the compatibility between the data and $H_0$:
\begin{equation}
  q_0 = \tilde{t}_{0} =
  \begin{cases}
    -2 \log \lambda(0) & \hat{\mu} \ge 0,\\
    0 & \hat{\mu} < 0.
  \end{cases}
  \label{eq:plr_q0_discovery}
\end{equation}

The level of disagreement with the background-only hypothesis is computed as
\begin{equation}
\begin{aligned}
  p_{0} &= P(\tilde{t}_0 \ge \tilde{t}_{0,\tx{obs}} | 0) 
  &= \int\limits_{\tilde{t}_{0,\tx{obs}}}^{\infty} f(\tilde{t}_{0} | 0) d\tilde{t}_{0},
\end{aligned}
  \label{eq:p_two_sided_disc}
\end{equation}
where $f(\tilde{t}_{0} | 0)$ is the probability distribution of $\tilde{t}_0$ under the assumption of the background-only hypothesis, $\mu=0$. The background-only hypothesis is rejected if $p_0$ falls below a predefined value, indicating that the data are not compatible with the no-signal hypothesis. 

\subsubsection{Discovery claims}
As is conventional in particle physics, this $p$-value can be expressed as a discovery significance in terms of the number of standard deviations $Z$ above the mean of a Gaussian that would result in an upper-tail probability equal to $p_{0}$, $\Phi^{-1}(1-p_{0})$, where $\Phi^{-1}$ is the inverse of the cumulative distribution function of the normal Gaussian distribution. In this formulation, a $3\sigma$ significance corresponds to a $p$-value of $p_{3\sigma}=1.4\times10^{-3}$ and a $5\sigma$ significance to a probability of $p_{5\sigma}=2.9\times10^{-7}$.

Following the convention in particle physics, we recommend that a global $p$-value smaller than $p_{3\sigma}$ is required for a claim of ``evidence.'' Section~\ref{subsec:lee} details the difference between the global $p$-value, which takes into account the effect of searching for several signals, and the local, $p_0$, which is computed only with reference to a fixed signal model. We recommend always reporting the smallest observed $p_0$ regardless of the presence or absence of any claim. Lastly, we recommend making available in supplementary material or upon request a plot of $p_0$ as a function of particle mass. We do not make a recommendation regarding the level of significance needed to claim ``discovery.''

\subsubsection{Look Elsewhere Effect}
\label{subsec:lee}
The ``look elsewhere effect'' (LEE) is a well-known phenomenon in searches for new physics \cite{Gross:2010qma} where, if testing the null hypothesis on the same set of data with respect to multiple alternatives, such as signal hypotheses featuring differing particle masses, the $p$-value needs to be corrected to account for the fact that a statistical fluctuation might be observed for any of the possible signal hypotheses\footnote{Technically, the condition is that one or more parameters of the signal hypothesis are degenerate under the null hypothesis}.  Failing to account for the LEE can lead to an overestimate in the apparent significance of a result. The size of the effect can be quantified by calculating the trial factor, the ratio of the $p$-value for observing an excess in a particular region to the global $p$-value for observing an excess for any of the signal-hypotheses. The size of this effect depends on the number of alternative models to the null hypothesis tested and the ability of the analysis to distinguish between them -- high-resolution peak searches will feature a large trial factor, while a counting experiment that cannot distinguish what signal model produces an excess will have a trial factor of 1. Therefore, the class of hypotheses considered when applying this correction should be included in reporting results. 

The LEE has not historically been evaluated in direct dark matter searches, except for a recent XENON1T publication~\cite{xenon1t_modstat}. To test the necessity of the LEE for dark matter searches, we follow the prescription of Ref~\cite{DundasMoraathesis}. Toy Monte-Carlos (MC) are generated and the discovery significance for every candidate mass $M$ is computed for each data set. 
The smallest local $p$-value for each toy data set, $p = \min_M(p_0(M))$, is recorded to estimate a ``probability distribution of smallest local $p$-values,'' called $f(p)$. The global significance $p_{\tx{data}}^{\tx{global}}$ for an observed excess with the smallest $p$-value, $p_{\tx{data}}$, is then:
\begin{equation}
    p_{\tx{data}}^{\tx{global}} = \int_0^{p_{\tx{data}}} f(p)dp .
\label{eq:lee_correction}
\end{equation}

\begin{figure}[tbp]
    \centering
    \includegraphics[width=0.9\columnwidth]{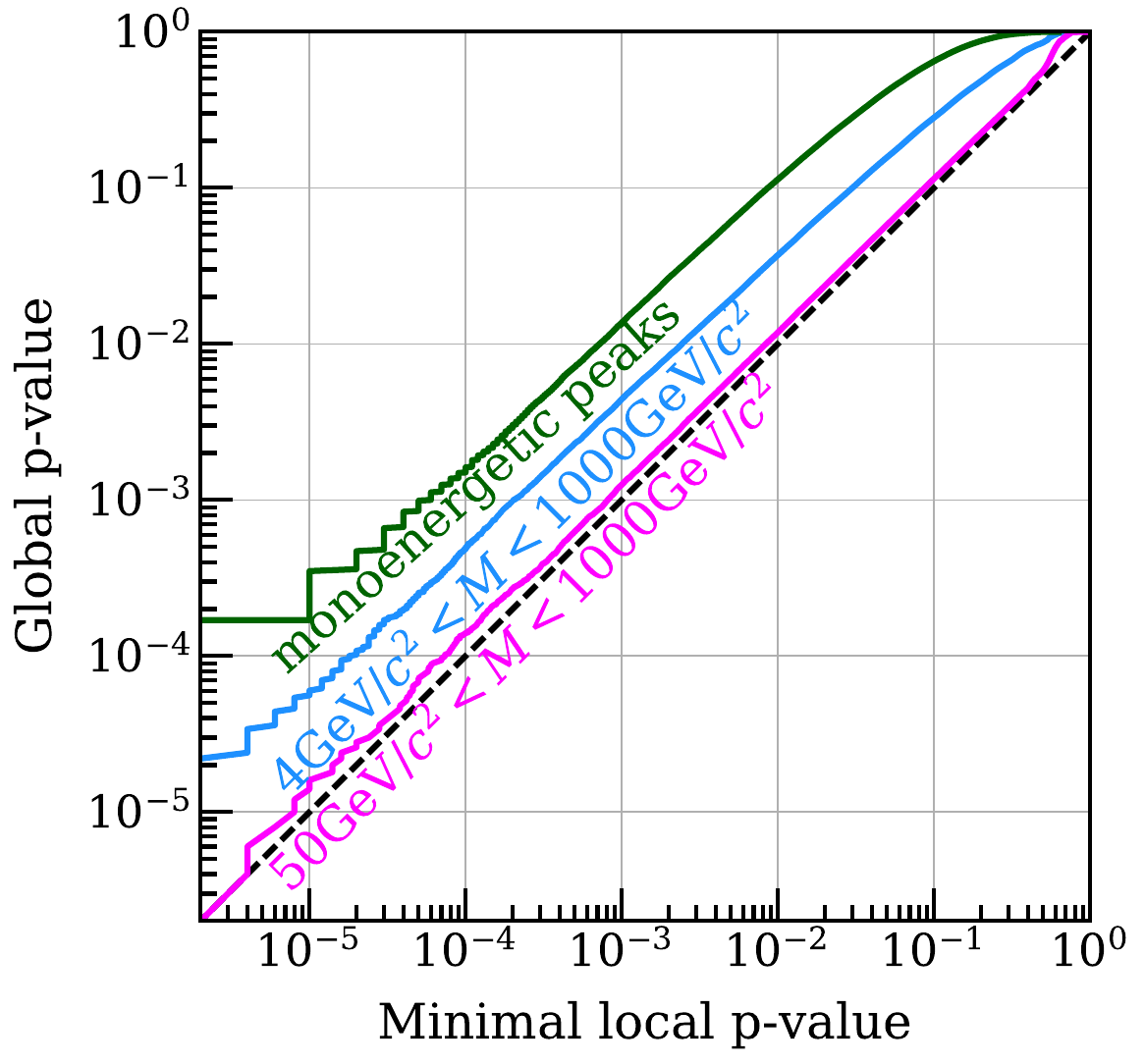}
    \caption{Illustration of the look elsewhere effect (LEE) for a search for dark matter with three families of dark matter models in a simplified model.
    Global $p$-values are plotted as a function of local $p$-value for a simplified LXe-TPC model computed with the procedure of Refs.~\cite{xenon1t_modstat, DundasMoraathesis}. A search for spin-independent recoils with WIMP masses between 50 $\gevm$ and 1000 $\gevm$ shows a negligible trial effect, as the spectra are almost degenerate. A search for spin-independent WIMPs with masses from 4 $\gevm$ to 1000 $\gevm$, representing a typical LXe-TPC mass range is shown in blue. The green line shows the trial factor for a search for 29 monoenergetic nuclear recoil peaks between 1 $\kev$ and 30 $\kev$.} 
    \label{fig:lookelsewhere}
\end{figure}

Figure~\ref{fig:lookelsewhere} shows the LEE evaluated for various types of searches in a simplified model of a LXe-TPC, demonstrating that the LEE can be significant when considering a range of dark matter masses and detector resolution typical for LXe-TPC searches. For searches restricted to masses above about 40 $\gevm$, the LEE is less important as the predicted recoil spectra are almost degenerate in the observable space. A search for monoenergetic peaks (such as an axion search), which effectively scans a range of statistically independent regions, leads to global $p$-values that are an order of magnitude greater than the minimum local $p$-value. Given the large computational cost associated with calculating the LEE, we propose that it be accounted for only if a local excess approaching or exceeding 3$\sigma$ is observed. If the computation needed to reach the relevant significance is unfeasible, alternative methods may be deployed if they can be shown to correctly account for the size of the effect (see for instance Ref.~\cite{Gross:2010qma}). 

\subsection{Limit Setting}
\label{sec:limits}
%
%

Confidence intervals and upper limits may be constructed via repeated hypothesis testing of a range of $\mu$ values, setting the endpoints of the interval (which may be one- or two-sided) at the critical point such that the $p$-value 
is equal to a predetermined value $\alpha$, also called ``the size of the test,'' or equivalently that the confidence level (CL) is $1-\alpha$.
Deciding which possible observations should be included in the confidence band for a certain true parameter value is referred to as choosing an ``ordering parameter,'' a test statistic with which to compute $p$-values that are used to define the confidence interval. This test statistic may or may not be the same test statistic that is used to compute discovery significance. Using the log-likelihood ratio as the test statistic to define the confidence interval yields the ``unified'' or Feldman-Cousins intervals~\cite{FeldmanCousins}, or, if there are nuisance parameters that are profiled over, the ``profile construction''~\cite{PDG, Cranmer2005}.
In this way, for a two-sided interval $[\mu_1, \mu_2]$,
\begin{equation}
1-\alpha \leq P(\mu_1 \le \mu_{\mathrm{true}} \le \mu_2). 
\label{eq:coverage}
\end{equation}
Here, the interval endpoints $\mu_1$ and  $\mu_2$ are random variables that depend on the experiment, with $\mu_{\mathrm{true}}$ the true, unknown value of the parameter of interest. A confidence interval method that fulfills Eq.~\ref{eq:coverage} for all possible signal hypotheses is said to have coverage -- a fraction $(1-\alpha)$ of the confidence intervals would contain the true value over repeated experiments.
For direct detection of dark matter, $\alpha$ is most often 0.1, leading to $90\%$ confidence levels, although a value of 0.05 ($95\%$ CL) is sometimes used. 

The two-sided test statistic most often used in dark matter limit setting is $\tilde t_\mu$ of Eq.~\ref{eq:plr_two_sided_mu_positive}. An alternative one-sided test statistic is:
\begin{equation}
  q_{\mu} =
  \begin{cases}
    -2 \log \lambda(\mu) & \hat{\mu} \le \mu,\\
    0 & \hat{\mu} > \mu,
  \end{cases}
  \label{eq:plr_one_sided}
\end{equation}
where $\lambda(\mu)$ is the profile likelihood ratio defined in Eq.~\ref{eq:plr}. With this definition, only the case $\hat{\mu} \le \mu$ is regarded as incompatible with the null hypothesis (see Ref.~\cite{CowanAsymptotic} for more details). 

\begin{figure}[tbp]
    \centering
    \includegraphics[width=0.9\columnwidth]{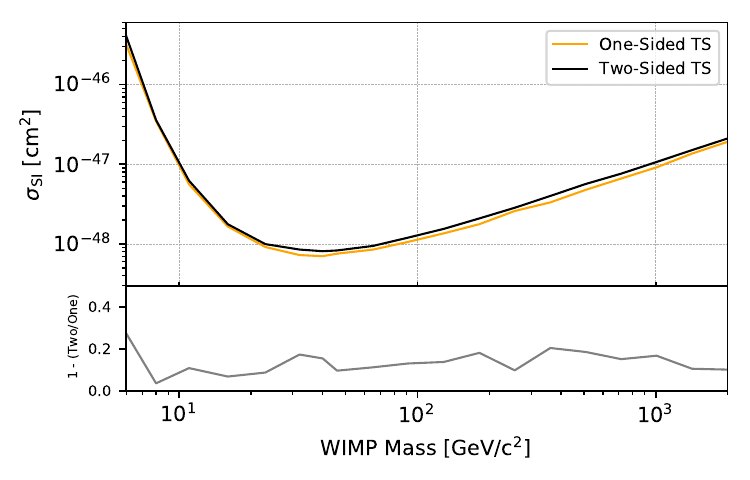}
    \caption{Comparison of the upper limit obtained using a one-sided (orange) and a two-sided (black) test statistic, respectively, for the same data set. The relative difference is indicated in the lower panel, and for this study, it can can be as large as 20\% for some masses.}
    \label{fig:1sidev2side}
\end{figure}

It is important to note that the choice of one-sided or two-sided test statistic can change the inferred result by a significant fraction for the same data set, as shown in Fig.~\ref{fig:1sidev2side}. Different direct dark matter experiments have used either the one- or two-sided PLR test statistic in their science papers (see, for instance, Refs.~\cite{xenon1t_modstat, pandax_54td, lux_complete}). Here, we recommend the two-sided construction of Eq.~\ref{eq:plr_two_sided_mu_positive}. This decision is motivated by the desire to use the same test statistic for limit setting as for discovery (recall that $q_0$ of Eq.~\ref{eq:plr_q0_discovery} is a limiting case of Eq.~\ref{eq:plr_two_sided_mu_positive}), with the only difference being the size of the test. If an excess is observed, the two-sided interval will naturally \dquotes{lift off} from a value of $\mu=0$, rejecting cases where $\mu$ is too small, while results compatible with the background still yield an upper limit. Using a single, unified Neyman construction~\cite{FeldmanCousins} that provides both these results as a function of the data avoids the potential of an experiment flip-flopping between several constructions\footnote{Flip-flopping is a term used
to refer to the fact that the coverage probability of a confidence interval may
be different to the nominal value if one makes an analysis choice, for example between a one- or
two-sided test \emph{after} looking at the data.}. Equation~\ref{eq:plr_two_sided_mu_positive} corresponds to the profile construction described by the Particle Data Group~\cite{PDG}, and, in the absence of nuisance parameters, is equivalent to that of Feldman\&Cousins~\cite{FeldmanCousins,CowanAsymptotic}. 
The cost of choosing the two-sided construction is a marginally weaker upper limit (see Fig.~\ref{fig:1sidev2side}). We argue that this is acceptable if the recommendation is widely adopted among dark matter collaborations, as no ``unfair'' advantage in the apparent limit can be gained by switching from two-sided to one-sided. We also note that assessing the viability of a particular physics model  in light of a published upper limit is subject to hidden uncertainties that dominate the difference between the two test statistics; in any case, such assessments should always be undertaken with caution. 

We recommend the use of MC techniques to construct
the test-statistic distributions (see Sec.~\ref{asymptotic}), as opposed to assuming that these distributions follow an asymptotic approximation. We also recommend performing coverage checks to show that the
actual coverage of the hypothesis test is similar to the nominal confidence, including if the true values of nuisance parameters differ from those assumed in the construction of the confidence interval; in the presence of nuisance parameters, coverage is not guaranteed, but practice has shown that it generally provides correct coverage.
In~\cite{xenon1t_modstat}, the coverage was checked with MC simulations assuming a different true nuisance parameter value than that assumed for the profile construction, investigating the robustness of the method to errors in the estimated nuisance parameters.

\begin{figure}[htbp]
    \centering
    \includegraphics[width=0.9\columnwidth]{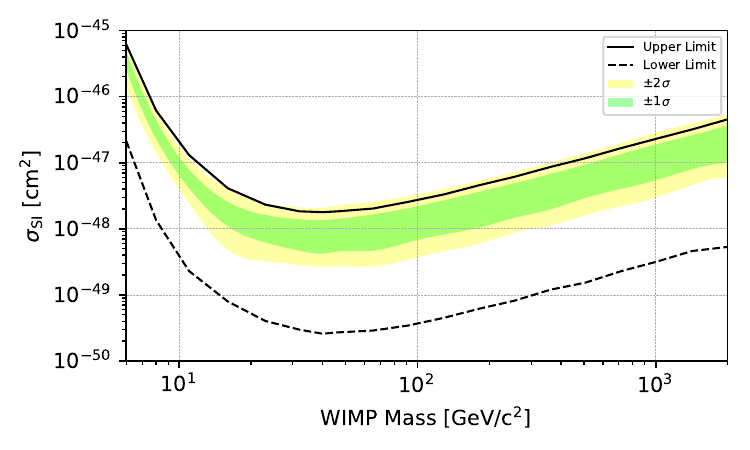}
    \includegraphics[width=0.9\columnwidth]{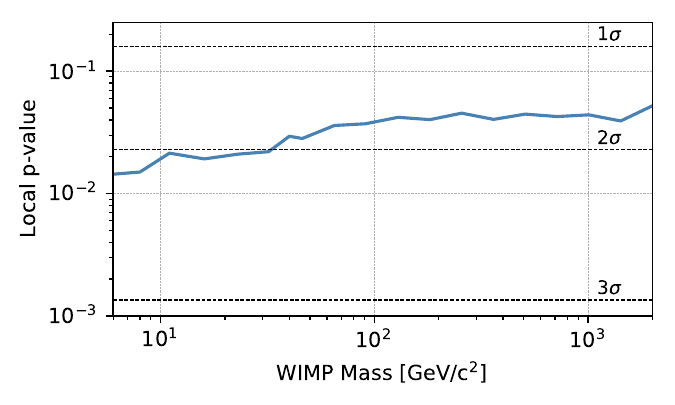}
    \caption{Top: Example two-sided $90\%$ CL limits (black) calculated from a single simulated background-only dataset, where the lower bound has ``lifted off'' from zero despite there being no signal in the data set. The green and yellow bands indicate bands containing $68\%$ and $95\%$ of upper limits under the null hypothesis.
    Bottom: The $p$-value for the background-only hypothesis as a function of mass. These data do not represent a statistically significant rejection of the background-only hypothesis.}
    \label{fig:liftingoff}
\end{figure}

Because limits are commonly set at $90\%$ CL, in the two-sided construction it is not unlikely that a data set will result in a non-zero lower limit on the parameter of interest despite not satisfying the requirement that the statistical significance is at least 3$\sigma$ to claim evidence of a positive signal. This is a natural consequence of frequentist hypothesis testing. As an example of such a case, the top panel of Figure~\ref{fig:liftingoff} shows $90\%$ CL upper and  lower limits from a hypothetical background-only experiment. Because $\alpha=0.10$, the backgrounds will fluctuate to give a lower bound in $10\%$ of cases. The lower panel presents the $p$-value versus WIMP mass, to show that these data do not approach a 3$\sigma$ significance. 

For a case like this, we recommend that collaborations should decide in advance on a significance threshold for reporting of a lower limit; for example, in the recent XENON1T publication~\cite{xenon1t_sr1}, if a result was less than 3$\sigma$ significant, no lower limit would be shown.  As stated previously, we recommend publication of the smallest observed $p$-value for the background-only hypothesis in addition to an upper limit in all cases, even if that $p$-value is not significant.  We also recommend collaborations publish the expected sensitivity of a result by showing a median expected limit with an uncertainty band (often called the "Brazil band").

\subsubsection{Cases with limited power}
\label{sec:limitedpower}
\begin{figure}[htbp]
    \centering
    \includegraphics[width=0.75\columnwidth]{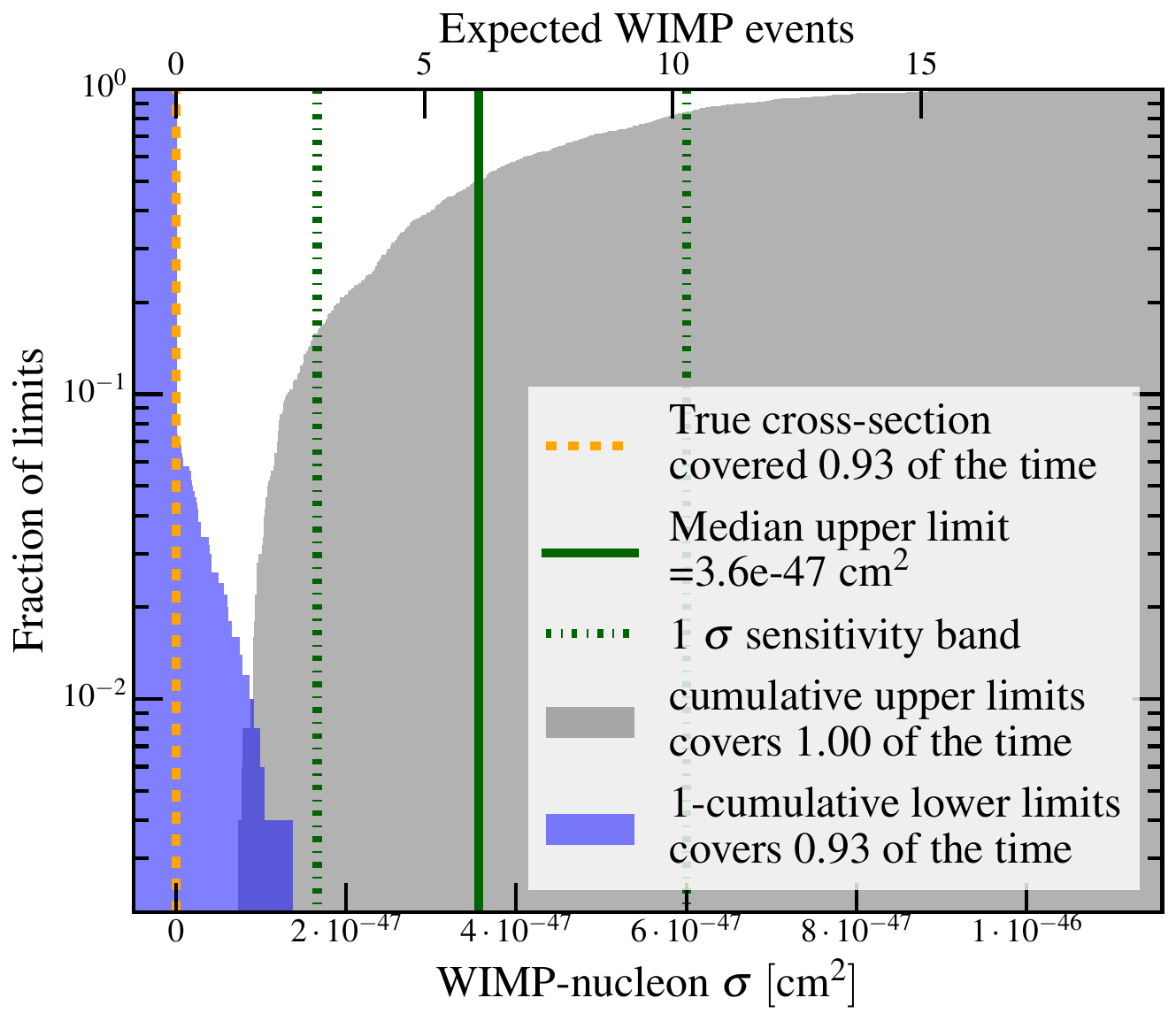}
    \caption{Expected distribution of upper and lower limits for XENON1T, for a $50~\gevm$ spin-independent WIMP search, with the true cross-section at $0$ (orange dashed line) from Ref.~\cite{xenon1t_modstat}.
    The fraction of upper limits excluding a certain cross-section represented by the x-axis are shown in gray; in other words, for an ensemble of trials with zero signal, what fraction of derived upper limits will exclude a given cross-section as a function of cross-section. The blue region shows the corresponding fraction for lower limits. The upper x-axis shows the signal expectation corresponding to the cross section on the lower x-axis in number of events---the upper limit tail reaches down to only two events.
    The dash-dotted green lines show the central $68\%$ of upper limits. The median upper limit is about a factor of $\sim$2 different from the 1$\sigma$ lines.}
    \label{fig:XENONULdistro}
\end{figure}

Sometimes confidence interval constructions may yield upper limits corresponding to signals much smaller than the ones to which the detector has any appreciable sensitivity or discovery power\footnote{In some cases an experiment can set a stronger limit in the presence of a (downwardly-fluctuating) background than an identical experiment with no background.}. In the case of an upper-limit only construction, this is purely an effect of the requirement to not cover even arbitrarily small signals a fraction $\alpha$ of the time.
As an example, Figure~\ref{fig:XENONULdistro} shows in gray an expected distribution of upper limits from the XENON1T experiment, with the distribution of upper limits extending to signal expectations of less than 2 events due to downward fluctuations of the background. 

A number of alternatives have been developed to address this concern~\cite{CLs,CLs2,PCL,cousins2011negatively},  
 and LUX~\cite{lux_complete}, PandaX-II~\cite{pandax_54td}, and XENON1T~\cite{xenon1t_sr1} have all at times applied the \dquotes{power-constrained limit} (PCL) of Ref.~\cite{PCL} to their upper limits, while the LHC community settled on the CLs construction of Refs.~\cite{CLs,CLs2}. Either of these constructions will cause overcoverage at very low signals, illustrated for example in Fig.~11 of~\cite{xenon1t_modstat}. 

\RevE{Here, we recommend applying the power constraint to limits obtained following Sec.~\ref{sec:limits}. We choose the PCL over the alternatives for its conceptual simplicity and because the CLs overcovers to higher quantiles. The principle behind power-constrained limits  is to use the rejection power of the experiment, $\pi({\mu})$ -- defined as the probability of rejecting a signal hypothesis $\mu$ through the upper limit when there is no true signal (i.e. $\mu$=0) -- as the metric to decide on the smallest signal that an experiment can exclude.\footnote{A previous version of this paper mistakenly defined the PCL threshold in terms of discovery power. Such a definition differs from both the original PCL paper and from how dark matter collaborations have used PCL. It also places little constraint on the limits of experiments with a small signal-background overlap, because these can discover signals on the observation of a single event in the signal region that is far from any background}
Setting a minimal rejection power, $\pi_{\mathrm{crit}}=\pi(\mu_\mathrm{crit})$, gives a minimal signal $\mu_\mathrm{crit}$. PCL was originally proposed for analyses that only set upper limits; although unified intervals show less severe downward fluctuations than classical upper limits~\cite{FeldmanCousins}, we would still apply PCL to constrain underfluctuations only. Therefore, we compute the exclusion power only considering upper limits excluding the signal even when using a two-sided construction.}

\RevE{A given rejection power corresponds to the same quantile of the distribution of background-only upper limits. For example, $\pi_\mathrm{crit} = 0.159$ would constrain limits to the -1$\sigma$ quantile of background-only upper limits, i.e. the lower green edge in the ``Brazil band.''} \RevE{The PCL method exactly preserves the original coverage of the interval construction for $\mu>\mu_\mathrm{crit}$. For the upper-limit-only construction proposed here, the PCL method gives a coverage of 1 for $\mu<\mu_\mathrm{crit}$. For unified intervals, it introduces less overcoverage, because it only affects the upper limit. }


The publication of Ref.~\cite{PCL} led to vigorous discussion on potential limitations of PCL in the literature and at various Phystat workshops, for example Ref.~\cite{cousins2011negatively}. A significant concern was whether increasing systematic uncertainties could lead to more stringent limits for certain choices of $\pi_\mathrm{crit}$ away from the median. For our purposes, an increase in systematic uncertainty not only widens the expected sensitivity bands of the limit but also raises (makes less sensitive) the median limit. The result is that both the median limit and, for example, the $-1\sigma$ band move up when a systematic is increased, and conservatism is maintained. For this reason, we feel comfortable moving forward with the PCL. 

The power threshold used in the power constraint is a fiducial choice in the analysis. A more conservative analysis might choose a higher threshold, such as the first LUX analysis~\cite{lux_first}, which demanded $\pi_\tx{crit}=0.5$. However, given the large random variation  in results of rare event searches (about a factor 2 around the median upper limit, see the difference between the median and 1$\sigma$ lines in green in Figure~\ref{fig:XENONULdistro}), this choice would somewhat arbitrarily limit the ability of  experiments to constrain a considerable swath of parameter space. The most recent publications by LUX, PandaX-II and XENON1T \RevE{constrained their limits to the -1$\sigma$ quantile} to maximize sensitivity while preserving the original purpose of the power constraint, \RevE{and we recommend using the -1$\sigma$ convention here. We further recommend collaborations be transparent about the use of any PCL, regardless of the choice of power, and the unconstrained limit should also be made available to the community.} 


\begin{figure}[htbp]
    \centering
    \includegraphics[width=0.95\columnwidth]{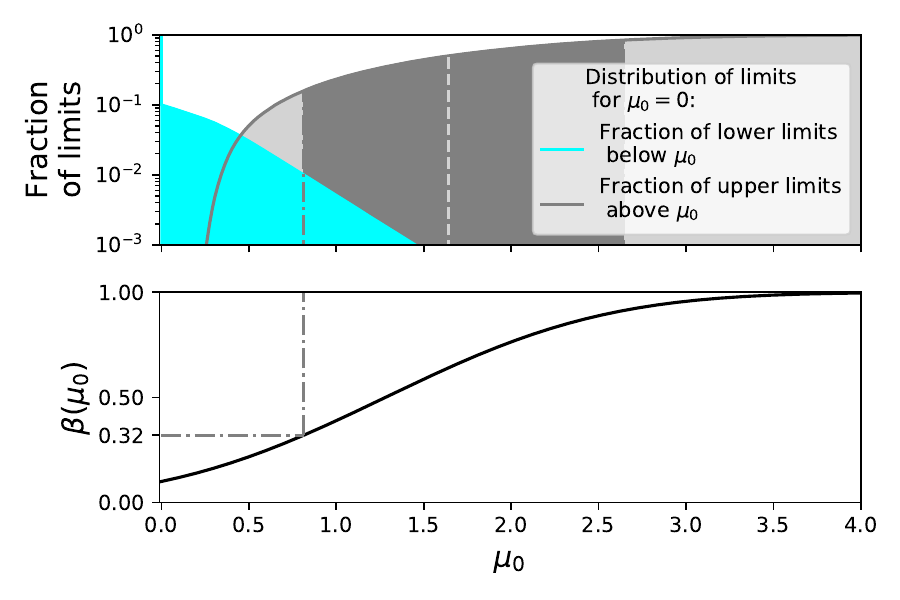}
    \caption{Distribution of upper and lower limits for a Feldman\&Cousins (FC) interval for a Gaussian measurement of a mean $\mu_0=0$ and $\sigma=1$ with a physical constraint of $0\leq\mu_0$, as an idealized analogue to Figure~\ref{fig:XENONULdistro}. The upper panel shows the fraction of signals excluded by an upper limit (gray) and lower limit (cyan), while the lower panel shows the power, $\pi(\mu_0)$, to discover a $p<0.1$ excess as a function of the signal. The dark gray band of the upper panel shows the central $68\%$ band of upper limits, typically shown as the sensitivity band by experiments. The dash-dotted line shows that for a $90\%$ FC construction and a Gaussian measurement, the lower edge of the sensitivity band corresponds to a power of $0.32$. (This figure is no longer referenced in the text after an Erratum was published. It is kept here to maintain consistency of figure numbers  across versions.)
    }
    \label{fig:gaussiancumulativeFC}
\end{figure}

\subsection{Asymptotic approximations}
\label{asymptotic}
\begin{figure}[tbp]
    \centering
    \includegraphics[width=0.8\columnwidth]{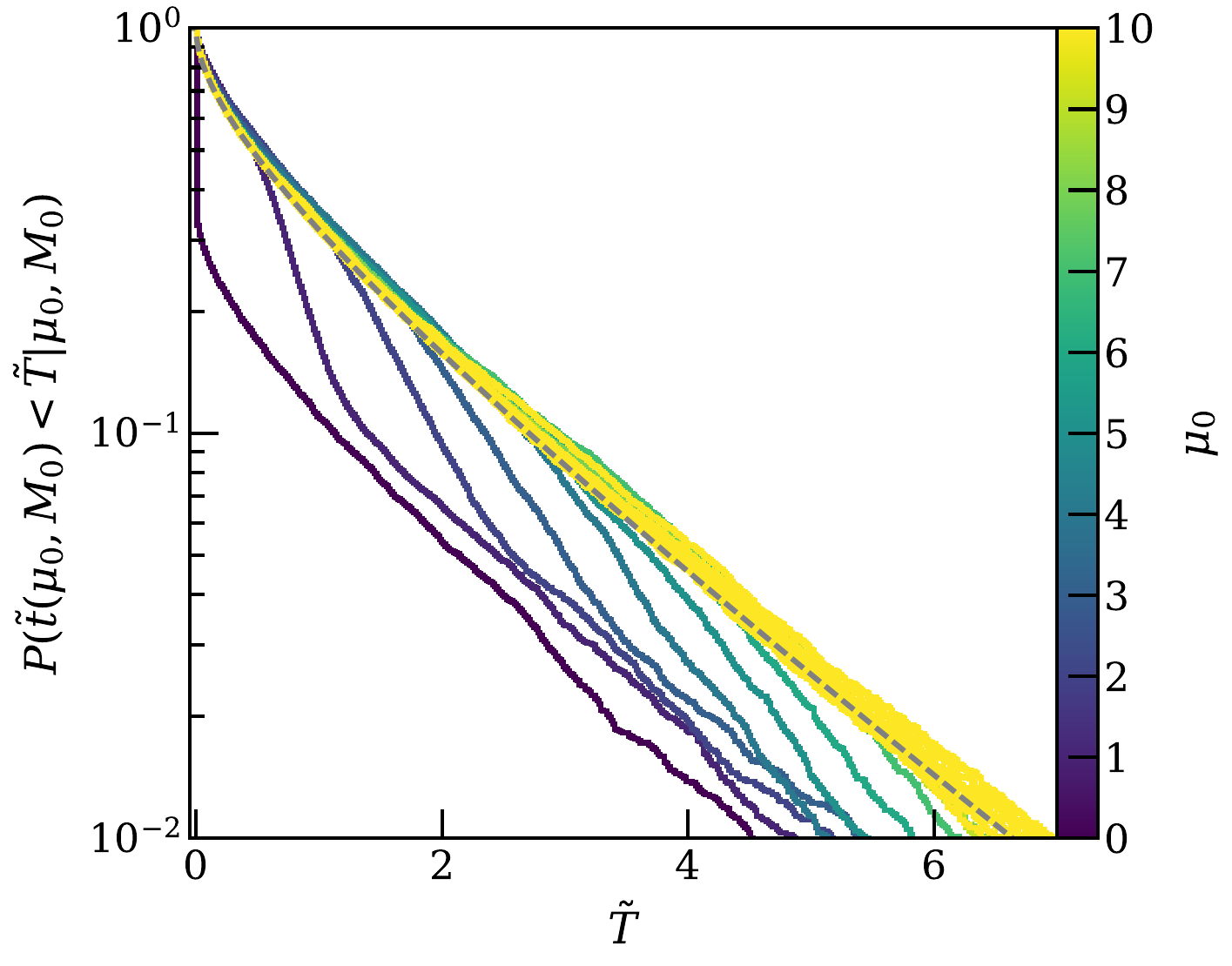}
    \caption{Probabilities estimated with toy-MCs for the test statistic $\tilde{t}(\mu_0, M_0)$ of equation~\ref{eq:plr_two_sided_mu_positive} to be smaller than a threshold value $\tilde{T}$ for a simplified LXe TPC likelihood, varying the true signal expectation $\mu_0$ between 0 and 10 events,
    for a $M_0=1000~\gevm$ SI WIMP. The dashed black line shows the asymptotic $\chi^2_1$ result for large signal expectations. 
    For small signal sizes (darker colored lines), the asymptotic approximation deviates significantly from the true $p$-value. 
    }
    \label{fig:asymptotic}
\end{figure}

Asymptotic formulae for test statistic distributions exist in the limit of infinite data~\cite{CowanAsymptotic}, and using the asymptotic approximation is a reasonable decision to save on computing time. 
In many cases, the approach to the asymptotic limit can be swift; for example, a counting experiment will reasonably approach the asymptotic result even for moderate expectation values ($\sim$5 events for $\alpha=0.1$). However, given the large background discrimination power in direct detection experiments, even results with hundreds of events may not converge to the asymptotic case because the expectation value in the signal region after discrimination is $\mathcal{O}(1)$ or less. 
Figure~\ref{fig:asymptotic} shows the distribution of a test statistic (solid colors) compared to an asymptotic approximation (dashed black) as the signal size increases for a simplified but representative simulation of a 1000 $\gevm$ dark matter search, similar to what is shown in Ref.~\cite{xenon1t_sr1}. For small numbers of signal events (darker colors), the asymptotic result poorly approximates the true test statistic distribution which is needed to compute discovery significances and confidence intervals.

If, as is often the case, toy MC simulations are used to estimate the distribution of the test statistic, a very significant result may require commensurately significant computational power to generate, for instance, the $> 10^7$ toy simulations needed to characterize a $5\sigma$ result. Nevertheless, we recommend that any usage of the asymptotic approximation be supported by MC studies to show its validity. In general, we recommend that sensitivity be calculated directly using adequate simulation, with the MC studies cross checked against uncertainty in the simulated values of the nuisance parameters. If a set procedure for this computation is in place, the actual simulation may need only be performed in the case that a highly significant result is seen. In the absence of adequate computing power to do full MC studies, arguments must be presented to justify whatever alternative methods are deployed.

\subsection{Contours in the Event of Discovery}
\label{contour}
In the discussion so far, we have assumed the hypothesis under test to be a signal model at a single dark matter mass. 
For excesses that approach discovery significance, however, collaborations may wish to perform parameter estimation of both the mass $M$ and cross section $\sigma$ in a vector-like parameter of interest $\bm{\mu}$ to form a 2D confidence contour. In such a case, we do recommend that collaborations set a significance threshold including the LEE effect before completing the analysis and removing any bias mitigation steps (see Sec.~\ref{sec:bias}) to determine whether a mass-cross-section contour should be included in a publication. The pre-determined threshold should be set high enough that flip-flopping between the per-mass cross sections and the 2D contour would introduce minimal bias (most likely satisfied by the requirement for a significant excess in the first place). Even if a 2D contour is reported, the per-mass confidence limit
should still be included. 

\subsection{Modeling Backgrounds and Detector Response}
\label{sec:backgrounds}

One of the requirements of the PLR method is that the model of detector response and backgrounds is correct.
Modelling these, and the validation thereof, is highly detector-specific and outside the scope of this paper. 
However, we believe it is essential for experiments to satisfactorily demonstrate goodness-of-fit for their background and detector models in order to properly utilize the methods presented here. 
This includes setting criteria for background model acceptance prior to an analysis, and clear presentation of those criteria in any eventual publication. 
One example of a goodness-of-fit criterion is the recent XENON publication, which required a background model $p$-value $\geq 0.05$ in a validation region in order to search for DM and solar $^8$B neutrino events in their data~\cite{Aprile:2020thb}. 
Whenever possible, models should be validated, both on calibration data or side-bands, and computing the goodness-of-fit of the best-fit model. 
The power of the goodness-of-fit test to detect impactful deviations from the assumed model should ideally be investigated. 
Uncertainties in the background model, when quantifiable, should be incorporated directly into the likelihood function as nuisance parameters and acknowledged as such in any publications. 

\subsection{Experimenter Bias Mitigation}
\label{sec:bias}
Experimenter bias is an effect which can, in general, drive a reported, measured value of a quantity away from its true value.  In this case, the choices that the analyzer makes regarding cuts and cut thresholds, analysis methods, and when to stop searching for errors in an analysis, are influenced by the quantitative result of the analysis.  Numerous examples in the historical physics literature have been identified in which new measurements of a physical quantity appear to be scattered around previous measurements, instead of being scattered around what we currently accept as the true values of those quantities\,\cite{Klein:2005di}.  

Methods to control for experimenter bias share a common approach: all choices an analyzer makes are taken without the analyzer knowing what effect those choices have on the final result.  In the case of DM experiments, four approaches have been employed, listed below.  We make no specific recommendations regarding bias mitigation, and leave such choices to the authors of a given result.
\begin{itemize}
    \item \emph{Signal blinding}: A plot is generally made in which observed events fall into various regions of parameter space characterized as more or less signal-like.  Often, DM experiments plot an electronic/nuclear recoil discriminant versus energy, and the low-energy ``nuclear-recoil-like'' area of the plot is considered to be the signal region.  In signal blinding, this region is masked for science data, but not for calibration data.  Only after all details of the analysis are frozen is this mask removed.  In this way, analysis details cannot be tuned based on the number of DM-like events that were observed.  The benefit of this type of bias mitigation strategy is that it is robust and simple to implement. The drawback is that rare backgrounds might exist in the data which will not be discovered until after the mask is removed.  Many examples of DM searches using signal blinding exist in the literature, including Refs.~\cite{Angle:2007uj, Aprile:2012nq, xenon1t_sr1, Lebedenko:2008gb, Agnese:2014aze,darkside_collaboration_darkside-50_2018}.
    \item \emph{Veto blinding}: A rare-event search such as a DM experiment will often entail the use of veto signals, which can identify when an event definitely does not result from the process under study.  Examples of such veto signals are ones which can tag cosmic rays in nearby materials, or acoustic sensors which can tag alpha decays in bubble chambers. If such a veto signal uniquely identifies background signals, one can choose to blind analyzers to that signal until all analysis details are finalized.  This provides analyzers a view of the signal region, but they are not able to know which events are signal and which are not.  The benefit of this type of approach is that analyzers have the opportunity to discover rare backgrounds because the signal region is viewable.  The drawback is that the background signals vetoed by such a tag may often not look quite like true signals, and therefore this technique may not be viable for some experiments.  Examples of this technique in use can be found in Refs.~\cite{Amole:2017dex, Aharmim:2011yq}.
    \item \emph{Salting}: An approach similar to that of veto blinding, salting is a technique where fake signal events are injected into the data stream. Analyzers may explore the signal region, but the identity, quantity, and distribution of these fake, injected events are kept blind to the analyzers.  The identities of the fake events are revealed only after the details of the analysis are finalized.  In this way, like veto blinding, this technique provides benefit of allowing the analyzer to identify rare backgrounds while being ignorant of the effect that analysis details have on the signal result.  The drawback to this approach is that it can be difficult to generate a collection of fake signal events that are believable.  The LIGO experiment has been able to inject fake gravitational waves by the use of hardware actuators~\cite{Biwer:2016oyg}; the LUX experiment constructed fake signal events from a sequestered calibration data set~\cite{lux_complete}.
    \item \emph{Data Scrambling}:
    An experiment may randomly smear data so that data in a control region and data in the signal region are randomly mixed. As an example, Antares introduced a random time-offset to each event when searching for neutrinos from dark matter in the Sun~\cite{antares_dm_2013}-- without removing this offset, it was impossible to determine if an event came from the Sun or another location on the sky. Similarly to salting, this allows all real events to be scrutinised before unblinding.
    
\end{itemize}

While one may, and often should, take steps to control for experimenter bias, it is important to note that this is not the only effect which can adversely influence the results of an analysis.  A holistic view, in which all systematic features are considered, is warranted.

\section{Astrophysical Models}
\label{sec:AP}
\subsection{WIMP Signal Model: Standard Halo Model}
\label{sec:Halo}

\begin{figure}[tbp]
    \centering
    \includegraphics[width=\linewidth]{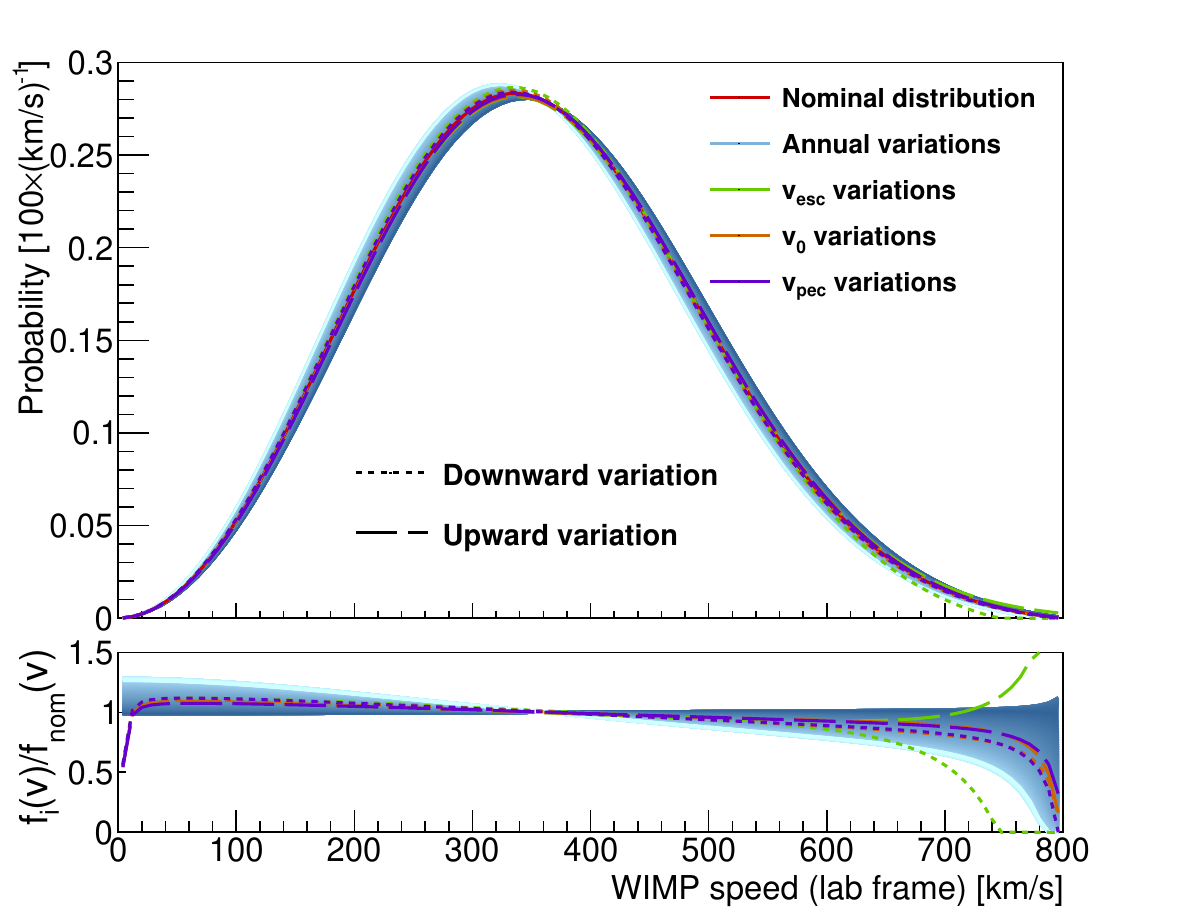}
    \caption{The Standard Halo Model WIMP speed distribution under varying parameter values. Variations throughout a year are shown in the blue gradient, with darker shades representing times closer to June. The red curve shows the speed distribution for the set of recommended parameters averaged over the full year, approximately equivalent to the distribution on March 9. Green and purple curves show the speed distributions with the galactic escape speed and the Sun's peculiar velocity ($v_\text{pec}=v_\varoast$) at their minimum and maximum values suggested by galactic survey analyses. The lower plot shows the ratio of each variation divided by the nominal model.}
    \label{fig:shm}
\end{figure}

The flux of WIMPs passing through the Earth is a necessary ingredient in the signal model for a WIMP hypothesis. Their galactic-frame velocity distribution, $f(\vec{v}_\text{gal})$, is usually assumed to be an isotropic Maxwell-Boltzmann distribution whose velocity dispersion $\sigma_0$ is defined by the local standard of rest at the location of the Sun, $|\vec{v}_0|=\sqrt{2}\sigma_0$, the Sun's peculiar velocity, $\vec{v}_\varoast$, and the Earth's velocity relative to the Sun, $\vec{v}_\varoplus$. Requiring that dark matter be gravitationally bound in the galaxy imposes an additional cut-off at the galactic escape speed, $v_\text{esc}$.
These assumptions result in a galactic-frame velocity distribution,
\begin{gather}
  \begin{aligned}
    f(\vec{v}_\text{gal}) &\propto \frac{\rho_\chi}{m_\chi}e^{-\frac{1}{2}|\vec{v}_\text{gal}|^2/\sigma_0^2}\Theta(v_\text{esc}-|\vec{v}_\text{gal}|), \\
    \vec{v}_\text{gal}    &= \vec{v}_\text{lab} + (\vec{v}_0+\vec{v}_\varoast+\vec{v}_\varoplus(t)),
  \label{eq:vdist}
  \end{aligned}
\end{gather}
where $\rho_\chi$ and $m_\chi$ are the local WIMP density and the WIMP mass, respectively, $\vec{v}_\text{lab}$ is the lab-frame WIMP velocity, and $\Theta(x)$ is the Heaviside step function. This ``Standard Halo Model'' (SHM) speed distribution is illustrated in the lab-frame in Figure~\ref{fig:shm}.

The time-dependent velocity of the Earth relative to the Sun is calculated in Refs.~\cite{mccabe_earthtextquotesingles_2014,ohare_velocity_2020}.
Defining the velocity vector as $(v_r,v_\phi,v_\theta)$, with $r$ pointing radially inward and $\phi$ in the direction of the Milky Way's rotation, this can be written as,
\begin{equation}
    \vec{v}_\varoplus(t) = \langle|\vec{v}_\varoplus|\rangle\times\left(
    \begin{array}{c}
         0.9941\cos(\omega\Delta t) - 0.0504\sin(\omega\Delta t) \\
         0.1088\cos(\omega\Delta t) + 0.4946\sin(\omega\Delta t) \\ 
         0.0042\cos(\omega\Delta t) - 0.8677\sin(\omega\Delta t) \\
    \end{array} \right),
    \label{eq:vearth}
\end{equation}
where $\omega=\SI{0.0172}{\day\tothe{-1}}$ and $\Delta t$ is the number of days since March 22, 2018 (an arbitrary date, and the choice of year has little effect). The average speed of the Earth is $\langle|\vec{v}_\varoplus|\rangle$, given in Table~\ref{tab:shmparams}.

The variation of the lab-frame WIMP speed due to the time evolution of $\vec{v}_\varoplus(t)$ is illustrated in Figure~\ref{fig:shm}. For most analyses not looking for annular modulation effects, it is sufficient to use the distribution averaged over the full year, which comes out to be approximately equivalent to the distribution evaluated at March 9,
\begin{equation}
    \vec{v}_\varoplus(\text{March 9}) = (\SI{29.2}{}, \SI{-0.1}{}, \SI{5.9}{})\SI{}{\km\per\second} .
    \label{eq:vearthave}
\end{equation}

With the exception of $m_\chi$, the parameters in Eq.~\ref{eq:vdist} constitute the SHM astrophysical parameters.
Since the model used to describe the flux of WIMPs influences the exclusion curves that are drawn, a unified approach to excluding WIMPs requires a consistent treatment of these parameters.
Recommended values for them are given in Table~\ref{tab:shmparams}.
The rationales for these parameter choices are discussed below.

Other authors have suggested updates to the SHM that differ from those presented here, including Refs.~\cite{radick_dependence_2020,evans_shm$^++$:_2018} among others. 
We recommend to report results with respect to the nominal halo model, which will give a common point of comparison, while knowledge of the dark matter halo continues to improve, unless new measurements significantly alter the expected spectra, particularly at high masses.  Shape variations in the velocity distribution can have an appreciable effect for limits on WIMPs that produce signals near the energy threshold of an experiment, but is otherwise not expected to have a major effect. 
Changes that affect the signal normalization but not the shape of the signal distribution, such as variations in the local dark matter density, can be easily accounted for by scaling published limits.

\begin{table*}[htb]
  \centering
  \caption{Suggested Standard Halo Model parameters. 
  Vectors are given as $(v_r,v_\phi,v_\theta)$ with $r$ pointing radially inward and $\phi$ in the direction of the Milky Way's rotation. 
  Analyses insensitive to annular modulation can approximate $\vec{v}_\varoplus(t)$ with Eq.~\ref{eq:vearthave}.}
  \label{tab:shmparams}
  \begin{tabular}{lllc}\hline\hline
    Parameter                         & Description & Value                                       & Reference \\\hline
    $\rho_\chi$                       & Local dark matter density & \SI{0.3}{\GeV\per\square c\per\cubic\cm}    & \cite{lewin_review_1996}                 \\
    $v_\text{esc}$                    & Galactic escape speed     & \SI{544}{\km\per\second}                    & \cite{smith_rave_2007}                   \\
    $\langle|\vec{v}_\varoplus|\rangle$ & Average galactocentric Earth speed     & \SI{29.8}{\km\per\second}                   & \cite{mccabe_earthtextquotesingles_2014} \\
    $\vec{v}_\varoast$                & Solar peculiar velocity & $(11.1, 12.2, 7.3)$\,\SI{}{\km\per\second}  & \cite{schonrich_local_2010}              \\
    $\vec{v}_0$                       & Local standard of rest velocity & $(0,238,0)$\,\SI{}{\km\per\second}          & \cite{bland-hawthorn_galaxy_2016,abuter_improved_2021}        \\
    \hline\hline
  \end{tabular}
\end{table*}


The SHM WIMP speed distribution is illustrated in Figure~\ref{fig:shm}, where the effects of varying the SHM parameters over the range of values motivated by galactic survey analyses are shown. In general, these effects tend to be comparable to or much smaller than the variation of the speed distribution over the course of a year.
The effects of varying these parameters on XENON1T's limits~\cite{Aprile:2017iyp} are explored in Ref.~\cite{wu_uncertainties_2019}.

Recent observations call into question the adequacy of the SHM itself, as evidence for several kinematically distinctive substructures have emerged from studies of data released by the \emph{Gaia} mission~\cite{gaia_collaboration_gaia_2018} and the Sloan Digital Sky Survey (SDSS)~\cite{york_sloan_2000}.
These substructures are likely the result of the Milky Way's formation history and merger events with other galaxies, and may include the \emph{Gaia} Sausage (or \emph{Gaia Enceladus})~\cite{evans_shm$^++$:_2018,necib_dark_2018,necib_inferred_2019,myeong_sausage_2018,bozorgnia_dark_2019}, among several others~\cite{myeong_discovery_2018,myeong_halo_2018,koppelman_one_2018,koppelman_characterization_2019,helmi_stellar_2008,ohare_velocity_2020,necib_evidence_2020}.
The effects of such substructures on direct detection experiments are demonstrated in Refs.~\cite{adhikari_constraints_2020,buch_implications_2020}.
Additionally, $N$-body simulations of the Large Magellanic Cloud indicate that its passage through the Milky Way could have produced a significant fraction of the local dark matter above the galactic escape speed~\cite{besla_highest-speed_2019}.
Due to these effects, quoted uncertainties on the SHM parameters do not accurately reflect the uncertainties in the dark matter halo, nor do they represent likelihood distributions that can be meaningfully profiled over.
The authors of this document therefore suggest that these parameters be fixed to clearly stated values, so that they can be reinterpreted under varying halo models.
We note that this is the approach followed by most collaborations in the field over the last decade. 

Most of the values suggested in Table~\ref{tab:shmparams} are consistent with those already in common use for WIMP direct detection experiments~\cite{lux_collaboration_results_2017,aprile_dark_2018,deap_collaboration_search_2019}. The most significant change suggested here is an updated value of $\vec{v}_0$. We emphasize here again that if these parameter values are adopted, the relevant references should always be cited.

\subsubsection{Local dark matter density: $\rho_\chi$}
Values for $\rho_\chi$ vary significantly between different measurements, typically in the range \SIrange{0.2}{0.6}{\GeV\per\square c\per\cubic\cm}.
The range of past and proposed measurements are best described in Refs.~\cite{read_local_2014, de_salas_dark_2020}.
This parameter normalizes the overall flux, but does not affect the predicted velocity distribution or the resulting WIMP-nucleon recoil spectra; as such, the total number of WIMP events expected in a direct detection experiment scales directly with $\rho_\chi$, and the net effect of changing its value is to linearly scale exclusion curves with the same factor by which $\rho_\chi$ changed.
Interpreting current limits in terms of different values of this parameter is therefore trivial, and the recommended value is the one most commonly used in direct detection experiments, as suggested by Ref.~\cite{lewin_review_1996}.

\subsubsection{Galactic escape speed: $v_\text{esc}$}
The galactic escape speed was measured by the RAVE survey~\cite{steinmetz2006radial} and later improved with the additions of SDSS~\cite{york_sloan_2000} and \emph{Gaia}~\cite{gaia_collaboration_gaia_2018} data. 
Measurements of $v_\text{esc}$ are summarized in Table~\ref{tab:vesc}. 
While some recent measurements seem to be trending towards somewhat lower values of $v_\text{esc}$, the values in Table~\ref{tab:vesc} are broadly consistent with each other and with a value around \SI{550}{\kilo\meter\per\second}. This value is also consistent with the value estimated in Ref.~\cite{koppelman_determination_2021}, using the \emph{Gaia} DR2 dataset. As such, the recommendation put forth in this document is to use $v_\text{esc}=\SI{544}{\kilo\meter\per\second}$ to maintain consistency with assumptions used for existing WIMP-nucleon cross section limits.

\begin{table*}[tbp]
  \centering
  \caption{Reported values of galactic escape speed. The measurement reported in~\cite{deason_local_2019}* is a re-analysis of the data set using the same priors used in~\cite{piffl_rave_2014}.}
  \label{tab:vesc}
  \begin{tabular}{lcccrcc}\hline\hline
    Year & Reference                & Survey      & Data release                          & C.L.              & $v_\text{esc}$ interval                                   & $v_\text{esc}$ median                           \\\hline
    2007 & \cite{smith_rave_2007}   & RAVE        & 1~\cite{steinmetz2006radial}          & \SI{90}{\percent} & \SIrange{498}{608}{\kilo\meter\per\second} & \SI{544}{\kilo\meter\per\second} \\
    2014 & \cite{piffl_rave_2014}   & RAVE        & 4~\cite{kordopatis2013radial}         & \SI{90}{\percent} & \SIrange{492}{587}{\kilo\meter\per\second} & \SI{533}{\kilo\meter\per\second} \\
    2017 & \cite{williams_run_2017} & SDSS        & 9~\cite{ahn_ninth_2012}               & \SI{68}{\percent} & \SIrange{491}{567}{\kilo\meter\per\second} & \SI{521}{\kilo\meter\per\second} \\
    2018 & \cite{monari_escape_2018}& \emph{Gaia} & 2~\cite{gaia_collaboration_gaia_2018} & \SI{68}{\percent} & \SIrange{517}{643}{\kilo\meter\per\second} & \SI{580}{\kilo\meter\per\second} \\
    2019 & \cite{deason_local_2019} & \emph{Gaia} & 2~\cite{gaia_collaboration_gaia_2018} & \SI{90}{\percent} & \SIrange{503}{552}{\kilo\meter\per\second} & \SI{528}{\kilo\meter\per\second} \\
    2019 & \cite{deason_local_2019}*& \emph{Gaia} & 2~\cite{gaia_collaboration_gaia_2018} & \SI{90}{\percent} & \SIrange{548}{612}{\kilo\meter\per\second} & \SI{580}{\kilo\meter\per\second} \\
    2021 & \cite{necib_substructure_2021} & \emph{Gaia} & 2~\cite{gaia_collaboration_gaia_2018} & \SI{68}{\percent} & \SIrange{477}{502}{\kilo\meter\per\second} & \SI{485}{\kilo\meter\per\second} \\
    \hline\hline
  \end{tabular}
\end{table*}

\subsubsection{Average galactocentric Earth speed: $\langle|\vec{v}_\varoplus|\rangle$}
In this document, we support the  use of
\begin{equation}
\langle|\vec{v}_\varoplus|\rangle=\,\SI{29.8}{\kilo\meter\per\second},
\end{equation}
as suggested in Ref.~\cite{mccabe_earthtextquotesingles_2014}, along with the accompanying time-evolving definition of $\vec{v}_\varoplus(t)$ approximately summarized in Eq.~\ref{eq:vearth}.
This value of $\langle|\vec{v}_\varoplus|\rangle$ is consistent with the one suggested in Ref.~\cite{lewin_review_1996}.

\subsubsection{Solar peculiar velocity: $\vec{v}_\varoast$}
The Sun's peculiar velocity was determined in~\cite{schonrich_local_2010}, by fitting data from the Geneva-Copenhagen Survey~\cite{holmberg_geneva-copenhagen_2009}.
Based on this analysis, the authors of Ref.~\cite{schonrich_local_2010} derive a peculiar velocity of 
\begin{equation}
    \vec{v}_\varoast=(11.1^{+0.69}_{-0.75}, 12.24^{+0.47}_{-0.47}, 7.25^{+0.37}_{-0.36})\,\SI{}{\kilo\meter\per\second},
\end{equation}
    with additional systematic uncertainties of $(1,2,0.5)$\SI{}{\kilo\meter\per\second}.
We support using this value in dark matter searches.

We note that the velocity in the galactic plane is faster than had been reported by previous measurements, based on an analysis of the \emph{Hipparcos} catalog~\cite{leeuwen_validation_2007}, which reported a value of $\vec{v}_\varoast=(10.00\pm0.36,5.25\pm0.62,7.17\pm0.38)$\SI{}{\kilo\meter\per\second}~\cite{dehnen_local_1998}. 
The decision to support the more recent measurement over the older one is based on the arguments in Ref.~\cite{schonrich_local_2010}.

\subsubsection{Local standard of rest velocity: $\vec{v}_0$}
%

In Ref.~\cite{reid_proper_2004}, the proper motion of Sagittarius A$^*$ was measured to high precision, implying that the angular velocity of the Sun around the center of the galaxy is given by
\begin{equation*}
  \frac{v_\varoast^\varodot+v_0^\varodot}{R_\varodot}=\SI{30.24\pm0.12}{\kilo\meter\per\second\per\kilo\parsec} ,
\end{equation*}
where $v_\varoast^\varodot$ and $v_0^\varodot$ give the components of $v_\varoast$ and $v_0$ in the galactic plane (the $\phi$ component), and $R_\varodot$ is the distance from the Sun to the galactic center.

Uncertainties in most previous estimates of $\vec{v}_0$ were driven by uncertainties in $R_\varodot$.
This distance was recently reported as $R_\varodot=$\SI{8275\pm9 (stat.) \pm 33 (syst.)} 
{\parsec} \cite{abuter_improved_2021}, implying $|\vec{v}_\varoast+\vec{v}_0|=$\SI{250.2\pm 1.4}{\kilo\meter\per\second}.

Combined with measurements of the Sun's peculiar velocity, $\vec{v}_\varoast$, this velocity implies that the local standard of rest has a speed of \SI{238.0\pm1.5}{\kilo\meter\per\second}.
We note that this velocity is consistent with the independently measured circular speed of \SI{240\pm8}{\kilo\meter\per\second} suggested in Ref.~\cite{reid_trigonometric_2014}, and the value \SI{229\pm11}{\kilo\meter\per\second} in Ref.~\cite{eilers_circular_2019}, albeit with smaller uncertainties.
Uncertainties in $v_0$ are driven by uncertainties in the Sun's peculiar velocity.

Previous limits on WIMP-nucleon cross sections used a value of \SI{220}{\kilo\meter\per\second}~\cite{lux_collaboration_results_2017,xenon1t_sr1,deap_collaboration_search_2019}, as suggested by Refs.~\cite{strigari_galactic_2013,green_astrophysical_2017,krauss_extracting_2018,koposov_constraining_2010}, which quote an uncertainty around \SI{\pm20}{\kilo\meter\per\second}.
We recommend updating this parameter to \SI{238}{\kilo\meter\per\second}; while this new value is within the uncertainty of the old one, the choice of this parameter and its smaller uncertainty can have a material impact on dark matter searches~\cite{mccabe_astrophysical_2010}.

\subsection{Astrophysical Neutrinos}
\label{sec:AP_nu}

Astrophysical neutrinos are expected to be an important background for the next generation of direct detection experiments. There are several sources of neutrinos arriving at Earth \cite{Vitagliano:2019yzm}, but not all of them are relevant for direct dark matter searches. In this section we outline the dominant neutrino background sources and make some recommendations that are pertinent to direct detection experiments.

\begin{figure}[tbp]
    \centering
    \includegraphics[width=\columnwidth]{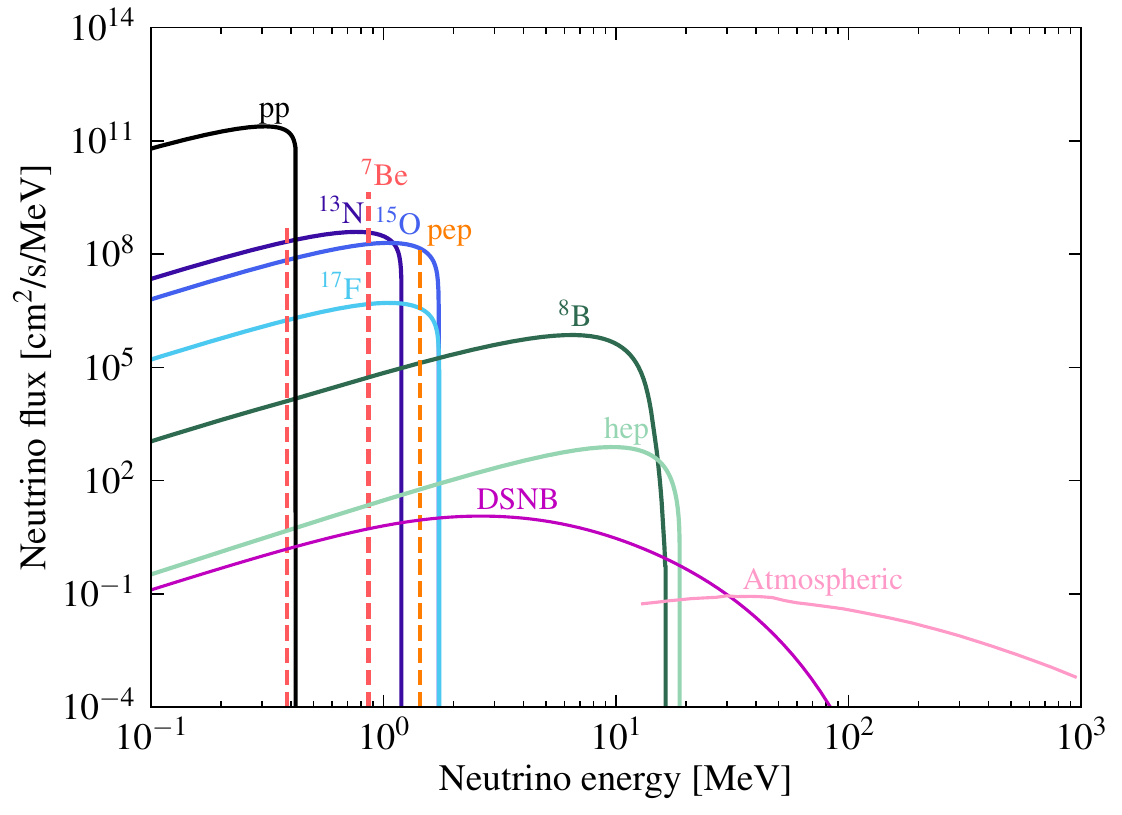}
    \caption{Dominant neutrino fluxes that constitute a background to direct detection experiments: solar, atmospheric, and DSNB. There are two monoenergetic $^{7}$Be lines at 0.38 and 0.86 MeV, indicated in red. The normalization values for each of these fluxes can be found in Table~\ref{tab:proposed_fluxes}.}
    \label{fig:nu_fluxes}
\end{figure}

Figure~\ref{fig:nu_fluxes} shows the neutrino fluxes that populate the relevant energy range for direct detection experiments. Low energy neutrinos from the \emph{pp} and $^{8}$Be solar reactions give rise to neutrino-electron scattering, which can become a prominent source of low energy electronic recoils. Nevertheless, the ultimate background might come from neutrino-induced nuclear recoils created by coherent neutrino-nucleus scattering, a process that has been recently confirmed experimentally by COHERENT \cite{Akimov:2017ade}. For example, in a xenon target $^{8}$B and \emph{hep} solar neutrinos can mimic a WIMP signal with a mass of approximately \SI{6}{GeV/c^2}, while atmospheric neutrinos and neutrinos from the diffuse supernova neutrino background (DSNB) will mimic a WIMP signal for masses above \SI{10}{GeV/c^2}. 
Next, we describe each of these neutrino sources separately.

\subsubsection{Solar neutrinos}
Our current understanding of the processes happening inside the Sun is best summarised by the Standard Solar Model (SSM), which originated more than three decades ago \cite{1963ApJ...137..344B}. According to the SSM, the Sun produces its energy by fusing protons into $^{4}$He via the $pp$ chain ($\sim$99\%) and the CNO cycle ($\sim$1\%). The SSM has been under constant revision since then, as more precise measurements and calculations of the solar surface composition and nuclear reaction rates become available. However, when modelling the solar interior based on a new generation of solar abundances \cite{2009ARA&A..47..481A}, the recent SSMs have failed to reproduce helioseismology data \cite{2004ApJ...606L..85B,Bahcall_2005,2006ApJ...649..529D,Serenelli:2009yc}; this is the so-called \dquotes{solar abundance problem}.  

The new generation of solar models are usually classified as high-Z and low-Z models, which reflect their different assumptions on the solar metallicity (Z). In the present work we adopt the most recent solar models developed in Ref.~\cite{Vinyoles:2016djt}. Figure~\ref{fig:nu_fluxes} shows the main contributions from the $pp$ chain and CNO cycle, and the overall normalization values are shown in Table~\ref{tab:solar_fluxes} for a high-Z (B16-GS98) and a low-Z (B16-AGSS09met) model, respectively. There is also a neutrino component arising from electron capture on the $^{13}$N, $^{15}$O and $^{17}$F nuclei \cite{Stonehill:2003zf,Villante:2014txa}, but their expected fluxes are very low. Since the CNO cycle has a strong dependence on the assumed metallicity, high-Z and low-Z models will predict different CNO neutrino fluxes. Also, a low-metallicity Sun will have a moderately cooler core, lowering the expected flux from the most temperature-sensitive neutrinos, such as those from the $^{8}$B and $^{7}$Be reactions \cite{Bahcall:1996vj,Robertson:2012ib}. Note that more precise measurements of the neutrino fluxes, by solar neutrino experiments or a next-generation liquid noble detector, will be crucial to resolve the solar abundance problem.

The photon luminosity of the Sun has been measured to a precision of less than 1\% \cite{doi:10.1029/1998GL900157}. The solar output is distributed into photon and neutrino channels, which introduces a direct constraint to the neutrino fluxes based on the measurement of the photon luminosity, most commonly known as the \dquotes{luminosity constraint} \cite{Bahcall:2001pf}. Since the $pp$ and $^{7}$Be reactions are dominant, their neutrino fluxes will also be dominant and the predicted uncertainty will be small to satisfy this constraint \cite{Smirnov2016,Bergstrom:2016cbh}. The CNO fluxes are also affected by this constraint, but on a smaller scale. For a recent discussion on this topic, see Ref.~\cite{Vescovi:2020wyz}.

\begin{table*}[tbp]
    \centering
    \caption{List of the solar neutrino fluxes that are relevant for direct dark matter searches. HZ and LZ stand for \dquotes{high-metallicity} and \dquotes{low-metallicity}, respectively. Values for the B16 solar models are from Ref.~\cite{Vinyoles:2016djt}.}
    \resizebox{0.95\linewidth}{!}{
    \begin{tabular}{lllll}
        \hline\hline
        \multirow{2}{*}{Name} & \multirow{2}{*}{End-point} & \multirow{2}{*}{B16-GS98 (HZ)} & \multirow{2}{*}{B16-AGSS09met (LZ)} & \multirow{2}{*}{Experimental} \\[2mm]
        & [MeV] & [cm$^{-2}$s$^{-1}$] & [cm$^{-2}$s$^{-1}$] & [cm$^{-2}$s$^{-1}$] \\
        \hline
        $pp$& 0.40 & $5.98(1\pm0.006)\times10^{10}$ & $6.03(1\pm0.005)\times10^{10}$ & $(6.1\pm0.5^{+0.3}_{-0.5})\times10^{10}$ \cite{PhysRevD.100.082004} \\
        $pep$ & 1.44 & $1.44(1\pm0.01)\times10^{8}$ & - & $(1.27\pm0.19^{+0.08}_{-0.12})\times10^{8}$ \cite{PhysRevD.100.082004} \\
        $pep$ & 1.44 & - & $1.46(1\pm0.009)\times10^{8}$ & $(1.39\pm0.19^{+0.08}_{-0.13})\times10^{8}$ \cite{PhysRevD.100.082004} \\
        $^{7}$Be & 0.38, 0.86 & $4.93(1\pm0.09)\times10^{9}$ & $4.50(1\pm0.06)\times10^{9}$ & $(4.99\pm0.11^{+0.06}_{-0.08})\times10^{9}$ \cite{PhysRevD.100.082004} \\ 
        $^{8}$B & 16.00 & $5.46(1\pm0.12)\times10^{6}$ & $4.50(1\pm0.12)\times10^{6}$ & $(5.25\pm0.16\pm0.12)\times10^{6}$ \cite{Aharmim:2011vm} \\
        $hep$ & 18.77 & $7.98(1\pm0.30)\times10^{3}$ & $8.25(1\pm0.30)\times10^{3}$ & $<2.3\times10^{4}$ (90\% CL) \cite{Aharmim_2006} \\
        $^{13}$N & 1.20 & $2.78(1\pm0.15)\times10^{8}$ & $2.04(1\pm0.14)\times10^{8}$ & \multirow{3}{*}{$7.0^{+2.9}_{-1.9}\times10^{8}$ \cite{Agostini:2020mfq}} \\
        $^{15}$O & 1.73 & $2.05(1\pm0.17)\times10^{8}$ & $1.44(1\pm0.16)\times10^{8}$ & \\
        $^{17}$F & 1.74 & $5.29(1\pm0.20)\times10^{6}$ & $3.26(1\pm0.18)\times10^{6}$ & \\
        \hline\hline
    \end{tabular}
    }
    \label{tab:solar_fluxes}
\end{table*}
Experimental measurements are also indicated in the last column of Table~\ref{tab:solar_fluxes}. These measurements are not entirely model-independent, and correlations between CNO and $pp$ chain neutrinos must be taken into account. There are two notable exceptions: the measurements of the $^{8}$B and $^{7}$Be neutrino fluxes. In the former case, the SNO experiment observed $^{8}$B neutrinos via three different reactions: neutral current (NC), charged current (CC), and elastic scattering (ES) \cite{Aharmim:2011vm}. Due to this favourable situation, the only theoretical input required for this analysis was the shape of the $^{8}$B energy spectrum, with the overall normalization being constrained by the NC measurement. In the latter case, the end-point energy of $^{7}$Be is well separated from all the known backgrounds and other neutrino signals, allowing for a measurement of this flux with an uncertainty below 3\% \cite{PhysRevD.100.082004}.

If a direct detection experiment were to take only the experimental values from Table~\ref{tab:solar_fluxes}, there would be a risk of adopting some measurements with overly large uncertainties, which are mainly driven by detector-specific effects. This could potentially be controlled by performing a global analysis that includes the likelihood from each of these neutrino experiments, albeit this might prove to be impractical. Similarly, the predictions from the solar models also present some problems and, as mentioned above, there is currently not one fully consistent solar model. Taking all this information into account, we recommend using the solar neutrino predictions described in Ref.~\cite{Vinyoles:2016djt}, except for the $^{8}$B and $^{7}$Be fluxes, for which we recommend adopting the experimental values due to their small uncertainty and independence from other neutrino signals. We believe this choice will provide the best sensitivity for direct detection experiments, while using a reasonable collection of flux uncertainties.

Furthermore, there are a few important ingredients that need to be taken into account when converting a neutrino flux into a recoil rate: neutrino oscillations \cite{Akhmedov_2010,Fukuda:1998mi}, the choice of form factor \cite{Helm:1956zz,Patton_2012}, electron binding effects \cite{Chen_2017}, and electroweak uncertainties \cite{PDG}, to name the main ones. We leave the particular considerations for each of these factors at the discretion of each collaboration. Also, we recommend using the prediction from the high-Z model presented in Table~\ref{tab:solar_fluxes}, except for those cases in which the difference in the expected event count between the two models is sufficiently large, in which case we recommend that the predictions from both models be reported. The level at which this difference is considered important is also left at the discretion of each collaboration, but the crucial point is that this comparison should be made in terms of expected counts at the detector under consideration.


\subsubsection{Atmospheric neutrinos}
Atmospheric neutrinos arise from the collision of cosmic rays in the atmosphere and the subsequent decay of mesons and muons. This neutrino flux spans a wide range of energies, and while the high-end ($>1$~GeV) has been well studied, the low energy region remains largely unexplored, which is the most relevant for dark matter searches. Currently, the best predictions on the atmospheric neutrino flux in the sub-GeV regime are based on the 2005 FLUKA simulations \cite{Battistoni:2005pd}. The sum of the predicted electron, anti-electron, muon and anti-muon neutrino fluxes from this simulation is shown in Fig.~\ref{fig:nu_fluxes}. At higher energies, we recommend adopting the more recent calculation of Honda et al \cite{Honda:2015fha}. 

The two main uncertainties associated with this flux at low energies are the uncertainty on the interaction cross section between cosmic rays and air nuclei, and the one arising from the Earth's geomagnetic field, which introduces a cut-off in the low end of the energy spectrum. Taking into account these two effects, the uncertainty on the atmospheric neutrino flux below \SI{100}{MeV} is approximately 20\% \cite{PhysRevD.83.123001,Okumura_2017}. It should be highligthed that the cut-off induced by the Earth's geomagnetic field is dependent on the detector's location, resulting in a larger atmospheric flux for detectors that are nearer to the poles \cite{PhysRevD.83.123001}. Our recommendation for the total flux and its uncertainty is shown in Table~\ref{tab:proposed_fluxes}.

\begin{table*}[tbp]
    \centering
    \caption{Recommended normalization values of all relevant neutrino fluxes for direct dark matter searches. This is a subset of the values shown in Table~\ref{tab:solar_fluxes}.}
    \begin{tabular}{llll}
        \hline\hline
        Name & Flux (theo.) [cm$^{-2}$s$^{-1}$] & Flux (exper.) [cm$^{-2}$s$^{-1}$] & Uncertainty \\
        \hline
        $pp$ & $5.98(1\pm0.006)\times10^{10}$ & - & 0.6\% \\
        $pep$ & $1.44(1\pm0.01)\times10^{8}$ & - & 1\% \\
        $^{7}$Be & - & $(4.99\pm0.11^{+0.06}_{-0.08})\times10^{9}$ & 3\% \\[1mm]
        $^{8}$B & - & $(5.25\pm0.16\pm0.12)\times10^{6}$ & 4\% \\[1mm]
        $hep$ & $7.98(1\pm0.30)\times10^{3}$ & - & 30\% \\
        $^{13}$N & $2.78(1\pm0.15)\times10^{8}$ & - & 15\% \\
        $^{15}$O & $2.05(1\pm0.17)\times10^{8}$ & - & 17\% \\
        $^{17}$F & $5.29(1\pm0.20)\times10^{6}$ & - & 20\% \\
        Atm. & $10.5\pm2.1$ & - & 20\% \\[1mm] 
        DSNB & $86\pm43$ & - & 50\% \\
        \hline\hline
    \end{tabular}
    \label{tab:proposed_fluxes}
\end{table*}

\subsubsection{Diffuse supernova neutrinos}
The diffuse supernova neutrino background (DSNB) refers to the cumulative flux of neutrinos from supernova explosions over the history of the Universe.
The expected total flux of the DSNB is not large compared to other neutrino sources, but it can be relevant for direct dark matter searches since it extends to a higher energy range than solar neutrinos. 

The neutrino spectrum of a core-collapse supernova is well-approximated by a Fermi-Dirac distribution, with temperatures in the range 3--8 MeV \cite{Keil:2002in}. The DSNB flux shown in Fig.~\ref{fig:nu_fluxes} assumes the following temperatures for each neutrino flavour: 3 and 5 MeV for electron and anti-electron neutrinos, respectively, and 8 MeV for the total contribution of the remaining neutrinos. For more details on this calculation, see Refs.~\cite{Strigari:2009bq,Beacom_2010}. There are some large theoretical uncertainties in this calculation, and therefore, following the recommendations from Refs.~\cite{Horiuchi:2008jz,Beacom_2010}, we recommend assigning an uncertainty of 50\% on the DSNB flux.

\section{Overall Recommendations}
\label{sec:summary}

We conclude by providing a list of the main recommendations from the sections above. 
These recommendations do not preclude the development of new methods if they can be shown to have appropriate statistical properties, as long as comparisons with previous results are reported in a transparent manner. 
Instead, this set of recommendations provide a common framework that will facilitate the comparison of results between different experiments.

\subsection{Statistical analysis}
\begin{itemize}
    \item We recommend collaborations to decide on all the choices covered in this paper before proceeding with final analyses, regardless of whether collaborations are employing bias mitigation techniques such as blinding or salting. Changes in the analysis due to, e.g. discovering bugs after an unblinding, should be pointed out when reporting results.
    \item We recommend PLR as the test statistic to use to assess discovery significance and to construct confidence intervals. Alternate methods should fulfill similar statistical properties, in particular of coverage.
    \item For standard WIMP searches we advocate performing these assessments on a per-mass basis, as in a raster scan.
    \item Discovery significance should be assessed with the discovery test statistic, Eq.~\ref{eq:plr_q0_discovery} (Eq.~\ref{eq:plr_two_sided_mu_positive} evaluated at $\mu=0$).
    
    \item If the signal hypothesis has free parameters not defined under the null hypothesis, a look-elsewhere-effect (LEE) computation should be performed to calculate a global significance, at least for a local significance that approaches or exceeds $3 \sigma$. 
      
    \item Claims of evidence should require at least a $3\sigma$ global discrepancy with the background-only hypothesis. We do not make a recommendation regarding discovery significance. 
  
    \item Experiments should publish their discovery $p$-value, both local and, if needed, global for any analysis.
    \item The unified confidence interval approach should be used to construct confidence intervals, using the two-sided test statistic of Eq.~\ref{eq:plr_two_sided_mu_positive}. Staying with past convention in the field, the primary limit should use $\alpha=0.1$ (i.e.~90\% CL). We recommend collaborations publish the expected sensitivity of a result by showing a median expected limit with an uncertainty band (often called the "Brazil band"). 
    
    \begin{itemize}
        \item The two-sided confidence interval will ``lift off'' from 0 signal when $p<\alpha$.  Collaborations may decide, before proceeding to     the final analysis, to apply an excess reporting     threshold to report the lower limit only above some greater significance level. Note, however, that this approach will in general lead to overcoverage. See Ref.~\cite{xenon1t_modstat} for a previous example by XENON1T.
            \end{itemize}
    
    \item To avoid large underfluctuations that would exclude parameter space \RevE{to which the experiment has little sensitivity}, 
    we advocate the use of a power-constrained limit (PCL) on the confidence intervals obtained using the test statistic of Eq.~\ref{eq:plr_two_sided_mu_positive}, with a power of at least $\pi_\mathrm{crit}=0.159$. This corresponds to \RevE{restricting limits to the -1$\sigma$ quantile of background-only upper limits.}
   \RevE{We recommend that collaborations make available the unconstrained limit in data releases, but leave it to collaborations to determine how best to present this information.}
    
    \item For excesses approaching or exceeeding $3\sigma$, a separate mass-cross-section confidence contour could be included. The complete procedure including this step would be: 
    \begin{itemize}
        \item Compute per-mass (local) discovery significance and per-mass confidence intervals---both of these should always be reported.
        \item If a local discovery significance indicates an excess, compute and apply the look-elsewhere-effect to report a global discovery significance.
        \item If the global discovery significance exceeds a pre-determined
        threshold, a separate mass-cross-section contour may also be included as part of reporting on the excess. The pre-determined threshold should be set high enough that flip-flopping between the per-mass cross-sections and the two-dimensional contour is less of a concern, and the per-mass confidence limit should still be included when reporting the result.
    \end{itemize}
    
    \item We recommend that the distribution of test statistics be estimated using either toy simulations or approximations (asymptotic or otherwise) verified using toy simulations.
    \item Whenever possible, models should be validated, both on calibration data or side-bands, using goodness-of-fit tests chosen to discover relevant model discrepancies. Tests and criteria should be decided before data is unblinded.

    \item We recommend that collaborations work to make their data more usable to the physics community than specific limits, by making results computer-readable and accessible by default, and by working to develop open statistical models/likelihoods for use by the community. 

    \item To avoid analysis biases, experiments should perform blind or salted analyses to the extent possible, committing to analysis and statistical conventions before studying the science data. 
\end{itemize}

\subsection{Astrophysical models}

\begin{itemize}
    \item The overall recommendation is to use the SHM parameters in Table~\ref{tab:shmparams}. The most significant is an updated value of $\vec{v}_0$, with all the other parameters being equal to the most commonly used values. We emphasize that if these parameter values are adopted, the relevant references should always be cited and citing this reference only is not sufficient.
    \item Due to non-parametric uncertainties in the form of the SHM itself, it is recommended not to profile over the SHM parameters' uncertainties. Instead, we recommend that these parameters are fixed to clearly stated values, so they can easily be reinterpreted under different halo models.
    \item The list of suggested normalization values for the relevant neutrino fluxes is shown in Table~\ref{tab:proposed_fluxes}. We recommend using the theoretical prediction for all the neutrino sources, except for $^{7}$Be and $^{8}$B, for which the most recent experimental values have a low uncertainty and are completely uncorrelated to other neutrino signals. 
    \item We leave at the discretion of each collaboration to make the choices that they consider most appropriate to convert neutrino fluxes into recoil rates. 
\end{itemize}

\section*{Acknowledgements}
The authors are indebted to Olaf Behnke, Louis Lyons, and Tarek Saab as co-organizers of the Phystat-DM workshop, and for subsequent comments and useful discussions on the recommendations found here. We would like to thank the Knut and Alice Wallenberg foundation for contributions to the Phystat-DM workshop. This material is based upon work supported by the U.S. Department of Energy, Office of Science, Office of High Energy Physics, under Award Number DE-SC0011702. DB is supported by the Kavli Institute for Cosmological Physics at The University of Chicago through an endowment from the Kavli Foundation. IB is grateful for the support of the Alexander Zaks Scholarship, The Buchmann Scholarship, and the Azrieli Foundation. JD is supported by the Science and Technologies Facilities Council (STFC) Grant No. ST/R003181/1. CM is supported by the Science and Technology Facilities Council (STFC) Grant ST/N004663/1. BvK is supported by the Deutsche Forschungsgemeinschaft (DFG, German Research Foundation) under the Emmy Noether Grant No. 420484612, and under Germany’s Excellence Strategy - EXC 2121 “Quantum Universe” – 390833306. MCP and DF are supported by the McDonald Institute.


\bibliographystyle{utphys}

\bibliography{bibliography}

\end{document}